\documentclass[fleqn,showkeys,reprint,aps,pra,onecolumn,superscriptaddress,showpacs,floatfix]{revtex4-2}
\usepackage[caption=false]{subfig}
\usepackage[T1]{fontenc}
\usepackage[latin9]{inputenc}
\usepackage{color}
\usepackage{textcomp}
\usepackage{graphicx}
\usepackage{subscript}
\usepackage{babel}
\usepackage{bbm}
\usepackage{tabularx}
\usepackage{multirow}
\usepackage{amsmath,amssymb,amsfonts,hyperref}
\usepackage{epsfig,slashed,placeins}
\usepackage{braket,latexsym,epstopdf,rotating}
\usepackage[export]{adjustbox}
\usepackage{xcolor}
\usepackage{array}
\begin{document}
\title{Precision calculation of the bound-electron \texorpdfstring{$g$}\textsc{} factor in molecular hydrogen ions}
\author{Ossama Kullie}
\affiliation{Theoretical Physics at Institute for Physics, \\ Department of Mathematics and Natural Science, University of Kassel, Germany}
%\email{kullie@uni-kassel.de}
%\thanks{Electronic mail:kullie@uni-kassel.de}
%%%%%%%%%%%%%%%%%%%%%%%%%%%%%%%%%%%%%%%%%%%%%%%%%%%
\author{Hugo D. Nogueira}
\affiliation{Laboratoire Kastler Brossel, Sorbonne Universit\'{e}, CNRS, ENS-Universit\'e PSL, Coll\`{e}ge de France, 4 place Jussieu, F-75005 Paris,France}
%%%%%%%%%%%%%%%%%%%%%%%%%%%%%%%%%%%%%%%%%%%%%%%%%%%
\author{Jean-Philippe Karr}
%%%%%%%%%%%%%%%%%%%%%%%%%%%%%%%%%%%%%%%%%%%%%%%%%%%
\affiliation{Laboratoire Kastler Brossel, Sorbonne Universit\'{e}, CNRS, ENS-Universit\'e PSL, Coll\`{e}ge de France, 4 place Jussieu, F-75005 Paris,France}
\affiliation{Universit\'{e} Evry Paris-Saclay, Boulevard Fran\c{c}ois Mitterrand, F-91000 Evry, France}
%%%%%%%%%%%%%%%%%%%%%%%%%%%%%%%%%%%%%%%%
\begin{sloppypar}
\begin{abstract}
We calculate the bound-electron $g$ factor for a wide range of rovibrational states of the molecular hydrogen ions ${\rm H}_2^+$ and HD$^+$. Relativistic and QED corrections of orders up to $\alpha^5$ are taken into account. All contributions are calculated in a nonrelativistic QED framework, except for relativistic corrections of order $(Z\alpha)^4$ and above, which are obtained by calculating the relativistic $g$ factor using a precise minmax finite element solution of the two-center Dirac equation. A relative accuracy of $4-5 \times 10^{-11}$ is achieved for the scalar $g$ factor component, which represents an improvement by more than three orders of magnitude over previous calculations. These results are useful for internal state identification and rovibraional spectroscopy of single molecular hydrogen ions in Penning traps, and open a new avenue towards precision tests of QED.
\end{abstract}
%%%%%%%%%%%%%%%%%%%%%%%%%%%%%%%%%%%%
\maketitle
\end{sloppypar}

%%%%%%%%%%%%%%%%%%%%%%%%%%%%%%%
\section{Introduction}\label{inttr}
%%%%%%%%%%%%%%%%%%%%%%%%%%%%%%%%%%%%%%%%%%%%%%%%%

There is currently considerable interest in accurately measuring rovibrational (RV) transition frequencies in molecular hydrogen ions (MHI) such as H$_2^+$~\cite{Schenkel2024,Doran2024} and HD$^+$~\cite{Alighanbari2020,Patra2020,Kortunov2021,Alighanbari2023} and comparing the values with predictions of ab initio theory~\cite{Korobov2017,Korobov2021}. 
Such comparisons allow for a series of applications such as determination of fundamental constants ~\cite{Karr2016,Karr2023,Mohr2025,Schiller2024}, tests of wave mechanics~\cite{Alighanbari2023} or Lorentz symmetry~\cite{Vargas2025}, and fifth-force searches~\cite{Germann2021,Alighanbari2023}.

Moreover, a new generation of experiments aiming at high-resolution spectroscopy of single MHI in Penning traps is under development. One of the perspectives of these experimental efforts is to perform spectroscopy of H$_2^+$ and its antimatter counterpart $\bar{\rm H}_2^-$ for improved tests of the CPT symmetry~\cite{Myers2018}. A key requirement for manipulation of a single molecular ion is the ability to detect its internal state in a nondestructive way, which was recently demonstrated on HD$^+$ via excitation of spin-flip transitions, exploiting the dependence of their frequencies on the internal state~\cite{Koenig2025}. This technique relies on predictions of the spin-flip frequencies, the precision of which is currently limited at the 0.1~ppm level by the theoretical uncertainty of the bound-electron $g$ factor~\cite{Hegstrom1979,Karr2021}.

While such precision is expected to be sufficient for state identification in most situations~\cite{Karr2021}, there are several motivations to improve further the calculation of the bound-electron $g$ factor in MHI. Firstly, for state identification, increased precision would be useful to resolve quasi-coincidences that may occur between spin-flip frequencies associated with different RV states. Secondly, in future Penning-trap RV spectroscopy experiments, the Zeeman shift will need to be accurately predicted and controlled. For example, in a 4-T magnetic field, the present 0.1~ppm uncertainty of the bound-electron $g$ factor translates to a $\sim 10$~kHz uncertainty on the Zeeman shift. This is significantly larger than the theoretical uncertainty of RV transition frequencies~\cite{Korobov2017,Korobov2021}, which would complicate the search for a transition. Finally, the experimental tools under development allow measuring the $g$ factor itself with high precision, offering a promising new route for tests of QED and determination of fundamental constants. In this perspective, it is obviously desirable to improve theoretical predictions of the $g$ factor.

In view of the low nuclear charge ($Z=1$), non relativistic quantum electrodynamics (NRQED) appears to be a particularly well suited approach to undertake this calculation. On the other hand, recent progress in high-precision numerical resolution of the two-center Dirac equation~\cite{Kullie2022,Nogueira2022,Nogueira2023,Kullie2024} opens the way to non-perturbative calculations (without expansion in $Z\alpha$). In the present work, we combine both approaches to calculate higher-order corrections to the $g$ factor of MHI, improving theoretical predictions by more than three orders of magnitude. Radiative corrections are evaluated in the NRQED framework, whereas next-to-leading-order relativistic corrections are obtained by calculating the relativistic $g$ factor and subtracting the leading-order contribution. For this term, relativistic wave functions are calculated by solving the Dirac equation for the electron in the field of two fixed point nuclei, using the highly accurate numerical minmax finite element method (FEM)\cite{Kullie20041,Kullie2022,Kullie2024}.

The paper is organized as follows. Theoretical expressions for contributions to the $g$ factor of leading order $\alpha^2$ up to $\alpha^5$ are summarized in Sec.~\ref{sec:corrections}. Then, in Sec.~\ref{sec:grel}, we present the derivation and numerical calculation of the relativistic $g$ factor in the Dirac framework. Calculations of all the other contributions are done in a nonrelativistic framework, and are presented in Sec.~\ref{sec:results} together with our final theoretical predictions.

%%%%%%%%%%%%%%%%%%%%%%%%%%%%%%%%%%%%%%%%%%%%%%%%%%%%%%%%%%%%%%%%%%%
\section{Relativistic and QED corrections to the \texorpdfstring{$g$ } \textsc{}factor} \label{sec:corrections}
%%%%%%%%%%%%%%%%%%%%%%%%%%%%%%%%%%%%%%%%%%%%%%
\subsection{NRQED expansion and notations}
%%%%%%%%%%%%%%%%%%%%%%%%%%%%%%%%%%%%%%%%%%%%%%

In the NRQED framework, corrections to the $g$ factor are expressed as an expansion in $\alpha$, $Z\alpha$, and $m_{\mathrm{e}}/M$, where $Z$ and $M$ are the charge and mass of a nucleus. Including only the terms that are considered in the present work, the bound-electron $g$ factor in HD$^+$ can be written as
\begin{equation}
g \left(v,N,M_N\right) = g{_\mathrm{e}} + \Delta g^{(2)} + \Delta g^{(3)} + \Delta g^{(4)} + \Delta g^{(5)} + \ldots \,, \label{eq:gfact-expansion}
\end{equation}
where $g_{\mathrm{e}}$ is the free-electron $g$ factor, and $\Delta g^{(n)}$ denotes a correction of leading order $\alpha^n$, which can be further separated into nonrecoil (i.e. zero-order in $m_{\mathrm{e}}/M$, $\Delta g^{(n)}_{\mathrm{nonrec}}$) and recoil ($\Delta g^{(n)}_{\mathrm{rec}}$) contributions. All terms of the right-hand side except for $g_{\mathrm{e}}$ depend on the RV and Zeeman state $\left(v,N,M_N\right)$, which is not explicitly indicated for brevity. Note that Eq.~(\ref{eq:gfact-expansion}) is not a pure expansion in $\alpha$, because its terms have prefactors that involve $g_{\mathrm{e}}$, which is itself given by an $\alpha$-expansion.

The dependence on the Zeeman quantum number in the above equation reflects the fact that the $g$ factor is anisotropic in molecular systems~\cite{Hegstrom1979,Karr2021}. More precisely, the sum of the correction terms in the r.h.s. can be expressed as the expectation value of an operator that is the sum of a scalar operator $G^{(0)}$ and a rank-2 operator $G^{(2)}$~\cite{Karr2021}. The $g$ factor of a given state can then be expressed as
\begin{equation} \label{eq:gM}
g \left(v,N,M_N\right) = g_s(v,N) + \frac{3M_N^2 - N(N+1)}{\sqrt{N(N+1)(2N-1)(2N+3)}} g_t(v,N) \,,
\end{equation}
where
\begin{equation}
g_s(v,N) = \frac{\left\langle v,N || G^{(0)} || v,N \right\rangle}{\sqrt{2N+1}} \,, \;\;\; g_t(v,N) = \frac{\left\langle v,N || G^{(2)} || v,N \right\rangle}{\sqrt{2N+1}} \,.
\end{equation}
Here and in Ref.~\cite{Karr2021}, a non-standard definition for the tensor operator $G^{(2)}$ is used: it is defined in such a way that its standard component $G^{(2)}_0$ coincides with $G_{zz}$. This differs from the standard definition by a global factor of $\sqrt{2/3}$~\cite{Karr2014}, which is taken into account in Eq.~(\ref{eq:gM}).

In the present work, several corrections to the $g$ factor are calculated in the adiabatic approximation. In this framework, the anisotropy is manifested by the fact that the $g$ factor depends on the orientation of the magnetic field with respect to the internuclear axis $z$. The $g$ tensor is then defined by writing the interaction of the electron spin with the magnetic field in the form~\cite{Hegstrom1979}
\begin{equation}
H_{\mathrm eff} = \frac{e}{2m_{\mathrm{e}}} \sum_{ij} g_{ij} s_e^i B^j \,. \label{eq:gtensordef}
\end{equation}
In view of the axial symmetry, it is sufficient to calculate the components $g_{\perp} = g_{xx} = g_{yy}$ and $g_{\parallel} = g_{zz}$ to fully determine the $g$ tensor. The scalar and tensor contributions $g_s$ and $g_t$ defined can be deduced from these quantities using the relationships~\cite{Karr2021,Ramsey1951}
\begin{subequations}
\begin{gather}
g_s = \frac{2}{3} g_{\perp} + \frac{1}{3} g_{\parallel} \,, \label{eq:link gs} \\
g_t =  \frac{2}{3} \sqrt{\frac{N(N+1)}{(2N-1)(2N+3)}} \left( g_{\perp} - g_{\parallel} \right) \,.
\label{eq:link gt}
\end{gather}
\end{subequations}

Atomic units ($e=\hbar=m_{\rm e}=1$, $c=1/\alpha$) are used throughout. Theoretical expressions are written in the case of a one-electron diatomic molecule with nuclear masses $m_1$, $m_2$ and charges $Z_1$, $Z_2$. $\mathbf{r}=(x,y,z)$ is the electron's position with respect to the geometrical center of the nuclei. In the adiabatic approximation, we use cylindrical coordinates: $\mathbf{r} = \rho \cos(\phi) \mathbf{e}_x + \rho \sin (\phi) \mathbf{e}_y + z \mathbf{e}_z$. The electron's positions with respect to both nuclei are denoted by $\mathbf{r}_1$, $\mathbf{r}_2$, and its momentum by $\mathbf{p}_e$.

%%%%%%%%%%%%%%%%%%%%%%%%%%%%%%%%%%%%%%%%%%%%%%
\subsection{Relativistic correction of order \texorpdfstring{$(Z\alpha)^2$} \textsc{} }
%%%%%%%%%%%%%%%%%%%%%%%%%%%%%%%%%%%%%%%%%%%%%%

The leading-order nonrecoil relativistic correction was first evaluated by Hegstrom~\cite{Hegstrom1979} in the adiabatic approximation. A more precise calculation that also included the recoil part was later performed in an exact three-body approach~\cite{Karr2021}. In both works the approximation $g_{\rm e} = 2$ was used; for clarity, we report here the expressions of both nonrecoil and complete three-body results with their exact dependence on $g_{\rm e}$. 
Note that the nonrecoil contribution is used for the separation of higher-order relativistic corrections, see Sec.~\ref{sec:grel} below.

%%%%%%%%%%%%%%%%%%%%%%%%%%%%%%%%%%%%%%%%%%%%%%
\subsubsection{Nonrecoil contributions} \label{sec-gHegstrom}
%%%%%%%%%%%%%%%%%%%%%%%%%%%%%%%%%%%%%%%%%%%%%%

In what follows, $\langle \rangle_{\rm ad}$ indicates that the expectation value is calculated in the adiabatic approximation.

The first term is an isotropic correction which can be interpreted as arising from the ``relativistic mass increase'' of the electron~\cite{Hegstrom1979}:
\begin{equation}
\Delta g^{(2)}_{s-{\rm nonrec}-A} = -\alpha^2 g_{\rm e} \frac{\left\langle v,N \right| \mathbf{p}_e^2 \left| v,N \right\rangle_{\rm ad}}{2}\,. \label{eq:delta-gs-A-ad}
\end{equation}

The other terms are anisotropic. A first one comes from the electronic spin-orbit Hamiltonian in the external field, due to to the magnetic-field-dependent part of the mechanical momentum:
\begin{subequations}
\begin{gather}
\Delta g^{(2)}_{\perp -{\rm nonrec}-B} = \frac{1}{2} \alpha^2 (g_{\rm e} - 1) \left\langle v,N \left| \left( \frac{\rho^2}{2} + z^2 \right) \left(\frac{Z_1}{r_1^3} + \frac{Z_2}{r_2^3} \right) \right| v,N \right\rangle_{\rm ad} , \\
\Delta g^{(2)}_{\parallel -{\rm nonrec}-B} = \frac{1}{2} \alpha^2 (g_{\rm e} - 1) \left\langle v,N \left| \rho^2 \left(\frac{Z_1}{r_1^3} + \frac{Z_2}{r_2^3} \right) \right| v,N \right\rangle_{\rm ad} .
\end{gather}
\end{subequations}
A second term comes from the second-order energy shift induced by the orbital part of the Zeeman Hamiltonian and the spin-orbit Hamiltonian. It only contributes to the perpendicular component:
\begin{equation}
\Delta g^{(2)}_{\perp -{\rm nonrec}-C} = \alpha^2 (g_{\rm e} - 1) \left\langle v,N \left| U_{\rm so} Q (E_0 - H_0)^{-1} Q U_{\rm Z}\right| v,N\right\rangle_{\rm ad} \,, \label{eq:dg2perpC}
\end{equation}
where
\begin{equation}
U_{\rm so} = \frac{ Z_1 \left( \mathbf{r}_1 \times \mathbf{p}_e \right)_x}{r_1^3} + \frac{ Z_2 \left( \mathbf{r}_2 \times \mathbf{p}_e \right)_x}{r_2^3} \,, \;\;\; 
U_{\rm Z} = (\mathbf{r} \times \mathbf{p}_e)_x \,,
\end{equation}
and $Q$ is a projection operator on a subspace orthogonal to $|v,N\rangle$. The total nonrecoil correction is
\begin{subequations}
\begin{gather}
\Delta g_{\perp-{\rm nonrec}}^{(2)} = \Delta g^{(2)}_{s-{\rm nonrec}-A} + \Delta g^{(2)}_{\perp -{\rm nonrec}-B} + \Delta g^{(2)}_{\perp -{\rm nonrec}-C} \,, \label{eq:g2perp-nonrec} \\
\Delta g_{\parallel-{\rm nonrec}}^{(2)} = \Delta g^{(2)}_{s-{\rm nonrec}-A} + \Delta g^{(2)}_{\parallel -{\rm nonrec}-B} \label{eq:g2par-nonrec}\,.
\end{gather}
\end{subequations}
The corresponding scalar and tensor contributions, $\Delta g_{s-{\rm nonrec}}^{(2)}$ and $\Delta g_{t-{\rm nonrec}}^{(2)}$, are obtained by applying Eqs.~(\ref{eq:link gs}-\ref{eq:link gt}).

%%%%%%%%%%%%%%%%%%%%%%%%%%%%%%%%%%%%%%%%%%%%%%
\subsubsection{Complete correction} \label{sec:g2tot}
%%%%%%%%%%%%%%%%%%%%%%%%%%%%%%%%%%%%%%%%%%%%%%

We now give the contributions including both nonrecoil and recoil terms, as derived in~\cite{Karr2021}. In doing so, we restore the exact prefactors depending on the free-electron $g$ factor, whereas in Ref.~\cite{Karr2021} expressions are written using the approximation $g_{\rm e} = 2$. Here, all expectation values are calculated in a full three-body framework using precise variational functions.

The first contribution is identical to $\Delta g^{(2)}_{s-{\rm nonrec}-A}$ given in Eq.~(\ref{eq:delta-gs-A-ad}), except that the calculation is done with three-body wavefunctions:
\begin{equation}
\Delta g^{(2)}_{s-{\rm nonrec}-A-{\rm 3b}} = -\alpha^2 g_{\rm e} \frac{\left\langle v,N \right| \mathbf{p}_e^2 \left| v,N \right\rangle}{2}\,.
\end{equation}
The second contribution is
\begin{subequations}
\begin{gather}
\Delta g_{s-B}^{(2)} = \frac {\left \langle v,N || \sigma^{(0)} || v,N \right \rangle}{\sqrt{2N+1}} \,, \\
\Delta g_{t-B}^{(2)} = \frac {\left \langle v,N || \sigma^{(2)} || v,N \right \rangle}{\sqrt{2N+1}} \,,
\end{gather}
\end{subequations}
with
\begin{align}
\sigma^{(0)} &= \frac{1}{3} \alpha^2 \left( \frac{c_1}{r_1} + \frac{c_2}{r_2} + \frac{c_{12}^{(1)} \mathbf{r}_1\!\cdot\!\mathbf{r}_2}{r_1^3} + + \frac{c_{12}^{(2)} \mathbf{r}_1\!\cdot\!\mathbf{r}_2}{r_2^3}\right) , \\
\sigma^{(2)} &= \frac{1}{6} \alpha^2 \left( \frac{c_1 Q_{11}^{(2)}}{r_1} + \frac{c_2 Q_{22}^{(2)}}{r_2} + \frac{c_{12}^{(1)} Q_{12}^{(2)}}{r_1^3} + + \frac{c_{12}^{(2)} Q_{12}^{(2)}}{r_2^3} \right) .
\end{align}
Here, $Q_{ab}^{(2)}$ ($a,b=1,2$) is the tensor having the Cartesian components
\begin{equation}
Q_{ab}^{(2)ij} = \mathbf{r}_a \!\cdot\! \mathbf{r}_b \delta_{ij} - 3 r_a^i r_b^j \,,
\end{equation}
and
\begin{subequations}
\begin{gather}
c_1 = \frac{1}{M^2} \left( ((g_{\rm e}-1)M - m_{\rm e}) m_1 Z_1 + m_1 m_{\rm e} Z_1 Z_2 - \frac{(g_{\rm e} M + m_1)(m_2 + m_{\rm e}) m_{\rm e} Z_1^2}{m_1} \right) , \\
c_2 = \frac{1}{M^2} \left( ((g_{\rm e}-1)M - m_{\rm e}) m_2 Z_2 + m_2 m_{\rm e} Z_1 Z_2 - \frac{(g_{\rm e} M + m_2)(m_1 + m_{\rm e}) m_{\rm e} Z_2^2}{m_2} \right) , \\
c_{12}^{(1)} = \frac{1}{M^2} \left( ((g_{\rm e}-1)M - m_{\rm e}) m_2 Z_1 - (m_1 + m_{\rm e}) m_{\rm e} Z_1 Z_2 + \frac{(g_{\rm e} M + m_1) m_2 m_{\rm e} Z_1^2}{m_1} \right), \\
c_{12}^{(2)} = \frac{1}{M^2} \left( ((g_{\rm e}-1)M - m_{\rm e}) m_1 Z_2 - (m_2 + m_{\rm e}) m_{\rm e} Z_1 Z_2 + \frac{(g_{\rm e} M + m_2) m_1 m_{\rm e} Z_2^2}{m_2} \right),
\end{gather}    
\end{subequations}
with $M=m_1+m_2+m_{\rm e}$. Finally, the contribution from the second-order shift induced by the Zeeman and spin-orbit interactions is
\begin{subequations}
\begin{gather}
\Delta g_{s-C}^{(2)} = \alpha^2 \frac {\left \langle v,N || T^{(0)} || v,N \right \rangle}{\sqrt{2N+1}} = \frac{\alpha^2}{3} \left( a_- + a_0 + a_+ \right), \\
\Delta g_{t-C}^{(2)} = \alpha^2 \frac {\left \langle v,N || T^{(2)} || v,N \right \rangle}{\sqrt{2N+1}} = \alpha^2\frac{\sqrt{N(N\!+\!1)(2N\!-\!1)(2N\!+\!3)}}{3} \left( -\! \frac{a_-}{N(2N\!-\!1)} \!+\! \frac{a_0}{N(N\!+\!1)} \!-\! \frac{a_+}{(N\!+\!1)(2N\!+\!3)} \right).
\end{gather}
\end{subequations}
Here, $a_-$, $a_0$, and $a_+$ are the contributions to the second-order perturbation term from intermediate states of angular momentum $N-1,N$, and $N+1$, respectively:
\begin{eqnarray}
a_- &=& -\frac{1}{2N+1} \sum_{n \neq 0} \frac{ \langle v N \| \mathbf{O}_Z^{(1)} \| v_n N-1 \rangle \langle v_n N-1 \| \mathbf{O}_{so}^{(1)} \| v N \rangle}{E_0 - E_n} ,\\
a_0 &=& \frac{1}{2N+1} \sum_{n \neq 0} \frac{ \langle v N \| \mathbf{O}_Z^{(1)} \| v_n N \rangle \langle v_n N \| \mathbf{O}_{so}^{(1)} \| v N \rangle}{E_0 - E_n} , \\
a_+ &=& -\frac{1}{2N+1} \sum_{n \neq 0} \frac{ \langle v N \| \mathbf{O}_Z^{(1)} \| v_n N+1 \rangle \langle v_n N+1 \| \mathbf{O}_{so}^{(1)} \| v N \rangle}{E_0 - E_n},
\end{eqnarray}
with
\begin{eqnarray}
\mathbf{O}_Z^{(1)} &=& \mathbf{L}_{eC} - \frac{Z_1 m_{\rm e}}{m_1} \mathbf{L}_{1C} - \frac{Z_2 m_{\rm e}}{m_2} \mathbf{L}_{2C} \,, \\
\mathbf{O}_{so}^{(1)} &=& \frac{1}{2} \left( \frac{Z_1}{r_1^3} (\mathbf{r_1} \!\times\! \mathbf{p}_e) + \frac{Z_2}{r_2^3} (\mathbf{r_2} \!\times\! \mathbf{p}_e) \right) - \left( \frac{m_{\rm e}}{m_1}\frac{Z_1}{r_1^3} (\mathbf{r_1} \!\times\! \mathbf{P_1}) +  \frac{m_{\rm e}}{m_2} \frac{Z_2}{r_2^3} (\mathbf{r_2} \!\times\! \mathbf{P_2}) \right).
\end{eqnarray}
$\mathbf{L}_{eC}$, $\mathbf{L}_{1C}$ and $\mathbf{L}_{2C}$ are the angular momenta of the electron and nuclei about the center of mass. The total relativistic correction including recoil terms is
\begin{subequations}
\begin{gather} \label{eq:gs2tot}
\Delta g_s^{(2)} = \Delta g^{(2)}_{s-{\rm nonrec}-A} + \Delta g^{(2)}_{s-B} + \Delta g^{(2)}_{s-C} \,,\\
\Delta g_t^{(2)} = \Delta g^{(2)}_{t-B} + \Delta g^{(2)}_{t-C} \,. \label{eq:gt2tot}
\end{gather}
\end{subequations}
The difference between the complete correction (calculated with precise three-body wavefunctions) and the nonrecoil terms given in the previous section (calculated in the adiabatic framework) corresponds to a sum of non-adiabatic and recoil effects, which we denote by $\Delta g^{(2)}_{\rm 3body}$:
\begin{subequations}
\begin{gather} \label{eq:gs3b}
\Delta g_{s-{\rm 3body}}^{(2)} = \Delta g_s^{(2)} - \Delta g_{s-{\rm nonrec}}^{(2)}\,,\\
\Delta g_{t-{\rm 3body}}^{(2)} = \Delta g_t^{(2)} - \Delta g_{t-{\rm nonrec}}^{(2)} \,. \label{eq:gt3b}
\end{gather}
\end{subequations}

%%%%%%%%%%%%%%%%%%%%%%%%%%%%%%%%%%%%%%%%%%%%%%
\subsection{Radiative correction of order \texorpdfstring{$\alpha(Z\alpha)^2$} \textsc{}} \label{sec:g3}
%%%%%%%%%%%%%%%%%%%%%%%%%%%%%%%%%%%%%%%%%%%%%%

The $\alpha(Z\alpha)^2$-order radiative correction originates from the electron's anomalous magnetic moment ($g_{\rm e} - 2$). Its expression (without recoil terms) was first derived in~\cite{Hegstrom1979} and can be confirmed using different variants of the NRQED approach~\cite{Pachucki2004,Pachucki2008,Kinoshita1996,Haidar2022}. In the present work, we evaluate this term numerically using the adiabatic approximation. It is expressed as follows:
\begin{subequations}
\begin{gather} \label{eq:g3perp}
\Delta g_{\perp - {\rm nonrec}}^{(3)} = \frac{1}{4} \alpha^2 (g_{\rm e} - 2) \left\langle v,N \left| \mathbf{p_e}^2 + p_{ez}^2 \right| v,N\right\rangle_{\rm ad} ,\\
\Delta g_{\parallel - {\rm nonrec}}^{(3)} = \frac{1}{2} \alpha^2 (g_{\rm e} - 2) \left\langle v,N \left| \mathbf{p_e}^2 - p_{ez}^2 \right| v,N\right\rangle_{\rm ad} . \label{eq:g3par}
\end{gather}
\end{subequations}

%%%%%%%%%%%%%%%%%%%%%%%%%%%%%%%%%%%%%%%%%%%%%%
\subsection{Higher-order relativistic corrections - the relativistic \texorpdfstring{$g$ }\textsc{} factor} \label{sec:grelho}
%%%%%%%%%%%%%%%%%%%%%%%%%%%%%%%%%%%%%%%%%%%%%%

The only contribution of leading order $\alpha^4$ is the pure $(Z\alpha)^4$ relativistic correction. This term could be derived in the NRQED framework, similarly to the $(Z\alpha)^2$ correction described above. However, in the present work, we take a different approach and calculate the  components $g^{\rm rel}_{\perp-{\rm nonrec}}$ and $g^{\rm rel}_{\parallel-{\rm nonrec}}$ of the relativistic $g$ factor, relying on precise numerical resolution of the two-center Dirac equation. Its calculation is presented in detail in Sec~\ref{sec:grel}. Since this quantity includes relativistic corrections to all orders in $Z\alpha$, the relativistic correction of orders $(Z\alpha)^4$ and above can be obtained by subtracting the leading-order terms from the result:
\begin{subequations}
\begin{gather} \label{eq:grel4perp}
\Delta g_{\perp - {\rm nonrec}}^{{\rm rel}(4+)} = g^{\rm rel}_{\perp-{\rm nonrec}} - 2 - g_{\perp - {\rm nonrec}}^{(2)} \,,\\
\Delta g_{\parallel - {\rm nonrec}}^{{\rm rel} (4+)} = g^{\rm rel}_{\parallel-{\rm nonrec}} - 2 - g_{\parallel - {\rm nonrec}}^{(2)}\,, \label{eq:grel4par}
\end{gather}
\end{subequations}
where the $(Z\alpha)^2$ terms in Eqs.~(\ref{eq:g2perp-nonrec}-\ref{eq:g2par-nonrec}) are calculated using the Dirac value $g_{\rm e} = 2$.

%%%%%%%%%%%%%%%%%%%%%%%%%%%%%%%%%%%%%%%%%%%%%%
\subsection{Radiative corrections of order \texorpdfstring{$\alpha(Z\alpha)^4$ } \textsc{}} \label{sec:g5}
%%%%%%%%%%%%%%%%%%%%%%%%%%%%%%%%%%%%%%%%%%%%%%

By far the largest correction at the $\alpha^5$ order is the one-loop self-energy, which was calculated in~\cite{Pachucki2004b} for hydrogenlike atoms. It comprises a logarithmic part (i.e. including an additional factor of $\ln ((Z\alpha)^{-2})$), which is a state-independent contribution proportional to the squared value of wavefunction at the electron-nucleus coalescence point (or delta-function expectation value). This term is thus readily obtained from the result of~\cite{Pachucki2004b}:
\begin{equation}
\Delta g_{s-{\rm SE}}^{(5)\ln} = \frac{32}{9} \alpha^5 \ln (\alpha^{-2}) \left\langle v,N \left| Z_1 \delta(\mathbf{r}_1) + Z_2 \delta(\mathbf{r}_2) \right| v,N \right\rangle.
\end{equation}
The nonlogarithmic part is state-dependent. In order to estimate this term, we approximate it by a delta-function potential. The coefficient chosen to match the result of~\cite{Pachucki2004b} for the 1S atomic state, which can be understood from the LCAO approximation, where the ground state electronic wavefunction is a linear combination of 1S atomic wavefunctions. This yields
\begin{equation} \label{eq:g5SE}
\Delta g_{s-{\rm SE}}^{(5)} \simeq a_{40}(1S) \alpha^5 \left\langle v,N \left| Z_1 \delta(\mathbf{r}_1) + Z_2 \delta(\mathbf{r}_2) \right| v,N \right\rangle,
\end{equation}
with $a_{40}(1S)=-10.236\, 524\, 318(1)$. Finally, the vacuum polarization correction at the same order, which was derived in~\cite{Karshenboim2000} for hydrogenlike atoms, is much smaller numerically. It is state-independent and is obtained from the result of~\cite{Karshenboim2000} as
\begin{equation}
\Delta g_{s-{\rm VP}}^{(5)} = -\frac{16}{15} \alpha^5 \left\langle v,N \left| Z_1 \delta(\mathbf{r}_1) + Z_2 \delta(\mathbf{r}_2) \right| v,N \right\rangle.
\end{equation}

To conclude this section, we recall that Eq.~(\ref{eq:gfact-expansion}) is not a pure expansion in $\alpha$, because of the $g_{\rm e}$-dependent prefactors. Whereas a systematic separation of orders in $\alpha$ could be obtained by expanding $g_{\rm e}$, we have opted for keeping the expressions in their current forms, which are simpler and have transparent physical origins.

We stress that this does cause any double counting issues. For example, expansion of $g_{\rm e}$ in the $(Z\alpha)^2$ contribution ($\Delta g_s^{(2)}$ and $\Delta g_t^{(2)}$, Eqs.~(\ref{eq:gs2tot}-\ref{eq:gt2tot})) produces a correction of order $\alpha(Z\alpha)^2$ that is distinct from the correction considered in Sec.~\ref{sec:g3}. It is worth noting that, in conjunction with the recoil part of the  $(Z\alpha)^2$ contribution, it also contains a part of the radiative-recoil contribution at orders $\alpha(Z\alpha)^2(m/M)^n$, which should be taken into account in a future complete calculation of this contribution. Finally, the calculation of relativistic corrections of order $(Z\alpha)^4$ and higher, as described in Sec.~\ref{sec:grelho}, is done with $g_{\rm e} = 2$, and thus does not include any $\alpha^n(Z\alpha)^4$ radiative corrections. Indeed, form factors corrections are fully included in the expressions of $\alpha(Z\alpha)^4$-order corrections given in Sec.~\ref{sec:g5}~\cite{Pachucki2004b}.

%%%%%%%%%%%%%%%%%%%%%%%%%%%%%%%%%%%%%%%%%%%%%%%%%%%%
\section{Calculation of the relativistic  \texorpdfstring{$g$} \textsc{} factor} \label{sec:grel}
%%%%%%%%%%%%%%%%%%%%%%%%%%%%%%%%%%%%%%%%%%%%%%%%%%%%

In this section, we derive and calculate numerically, in the Dirac framework, the components of the relativistic $g$ tensor, $g^{\rm rel}_{\perp - {\rm nonrec}}$ and $g^{\rm rel}_{\parallel - {\rm nonrec}}$ (see Sec.~\ref{sec:grelho}). In order to alleviate the notations, throughout this section they will be simply denoted by $g_{\perp}$ and $g_{\parallel}$.

%%%%%%%%%%%%%%%%%%%%%%%%%%%%%%%%%%%%%%%%%%%%%%
\subsection{Interaction with a magnetic field}
%%%%%%%%%%%%%%%%%%%%%%%%%%%%%%%%%%%%%%%%%%%%%%

The relativistic Hamiltonian that describes the interaction of an electron with an external magnetic field $\mathbf{B} = (B_x,B_y,B_z)$ can be written as
\begin{equation}\label{Veq}
H_{\rm int}=
-\frac{c}{2}
{\boldsymbol{\alpha}} \cdot \left( {\mathbf r}  \times \mathbf{B} \right) ,
\end{equation}
where ${\boldsymbol{\alpha}}=(\alpha_x,\alpha_y,\alpha_z) $ are the Dirac matrices. For the case of a field parallel to the internuclear axis $\mathbf{B}_{||}=(0,0,B_z)$ we obtain 
\begin{eqnarray}\label{Hpar}
 {H}_{||} &=&
\frac{c}{2} B_{\rm z}
\begin{bmatrix} 
  0 & 0 & 0& -i\rho\,e^{-i \phi}\\ 
  0 & 0 & i\rho\,e^{i \phi} &0 \\
  0& -i\rho\,e^{-i \phi} & 0&0 \\
  i\rho\,e^{i \phi}&0 & 0&0 
\end{bmatrix} \label{PaMat1}\\
&\equiv& \frac{c}{2}\,B_{\rm z}\, M_{||}(\rho,\phi) \nonumber \,.
\end{eqnarray}
If the field is perpendicular to the internuclear axis, e.g. along $x$, $\mathbf{B}_{\perp}=(B_x,0,0)$, one gets
\begin{eqnarray}\label{Hperpx}
 {H}_{\perp}^{\rm x} &=&
 -\frac{c}{2}\,  B_{\rm x}\, \begin{bmatrix} 
  0 & 0 &-\rho\, \sin\phi&  -i\,z \\
  0 & 0 & i\,z &\rho\, \sin\phi\\
  -\rho\, \sin\phi&  -i\,z & 0&0 \\
  i\,z  & \rho\, \sin\phi &0 & 0
\end{bmatrix}
\\\nonumber %\label{PeMat2}
&\equiv &-\frac{c}{2}\, B_{\rm x}  (M_{\perp}(z)+M_{\perp} (\rho,\sin(\phi)) )
\equiv H_{\perp}^{(1)}+ H_{\perp}^{(2)}.
\end{eqnarray}

%%%%%%%%%%%%%%%%%%%%%%%%%%%%%%%%%%%%%%%%%%%%%%%%%%%%%%%%%
\subsection{Components of the relativistic \texorpdfstring{$g$} \textsc{} tensor}  
%%%%%%%%%%%%%%%%%%%%%%%%%

Due to the axial symmetry, the solutions of the two-center Dirac problem can be classified with respect to the quantum number $m = \langle j_z \rangle$, where $j_z$ is the projection of the angular momentum on the internuclear axis $z$. Here, we focus on the ground $1s\sigma$ electronic level, which contains two degenerate states $m=\pm1/2$. For a $m=+1/2$ state, the angular dependence can be separated as follows~\cite{Mueller1976,Kullie2022}:
\begin{eqnarray}\label{eq:sep-angle}
  |\Psi_{+}\rangle =
 \begin{bmatrix}
 \psi_1& \\
 \psi_2\, e^{i \phi}& \\
 i \psi_3&\\
 i \psi_4\, e^{i \phi}
 \end{bmatrix},
\end{eqnarray}
where $\Psi_{+}$ stands for $\Psi_{m=+1/2}$, and $\psi_i, i=1-4$  are real functions of radial variables. Applying the time reversal operator~\cite{Sakurai2020}
\begin{equation}
\mathcal{\widehat{T}}=-i\, 
\begin{pmatrix}
\sigma_y & 0 \\
0 & \sigma_y 
\end{pmatrix} \, \mathcal{\widehat{C}},
\end{equation}
where $\mathcal{\widehat{C}}$ is the complex conjugation operator (note that $\mathcal{\widehat{T}}$ is defined up to some arbitrary phase factor, here we make a choice that we find convenient), we obtain the wave function $\Psi_{m=-1/2} \equiv \Psi_-$ .
\begin{eqnarray}\label{eq:timerevWF}
  |\Psi_{-}\rangle =\mathcal{\widehat{T}} |\Psi_{+}\rangle= 
  \begin{bmatrix}
 -\psi_2\, e^{-i \phi}& \\
  \psi_1& \\
 i \psi_4\, e^{-i \phi}\\ 
-i \psi_3&\\
 \end{bmatrix}.
 \end{eqnarray}
According to perturbation theory for degenerate states, the first-order energy shifts are found by diagonalizing the matrix
\begin{eqnarray} \label{eq:Hintmatrix}
 H_{int}^{(0)}=
  \begin{bmatrix}
 \langle\Psi_{+}| H_{int} |\Psi_{+}\rangle& \langle\Psi_{+}| H_{int}|\Psi_{-}\rangle\\
\langle\Psi_{-}| H_{int} |\Psi_{+}\rangle  & \langle\Psi_{-}| H_{int} |\Psi_{-}\rangle
 \end{bmatrix}.
 \end{eqnarray}
 
%%%%%%%%%%%%%%%%%%%%%%%%%%%%%%%%%%%%%%%%%%%%%%%%%%%%%%%%%%%%%
\paragraph{Parallel field.}
%%%%%%%%%%%%%%%%%%%%%%%%%%%%%%%%%%%%%%%%%%%%%%%%%%%%%%%%%%%%%

For the case of a field oriented along the internuclear axis $z$, $H_{int}$  is given in Eq. \ref{Hpar}. The diagonal elements are (see Appendix \ref{apA1} for details):
\begin{eqnarray}
    \langle\Psi_{+}|  H_{||} |\Psi_{+}\rangle
    %\\\nonumber
    &=&\frac{c}{2} B_{\rm z}  \langle\Psi_{+}| M_{||}(\rho,\phi) |\Psi_{+}\rangle \nonumber \\
    &=& 2\pi \, c\, B_{\rm z} \left(\langle\psi_{1}| \rho |\psi_{4}\rangle_{\rm rad}
    -\langle\psi_{2}| \rho |\psi_{3}\rangle_{\rm rad} \right), \label{gpar}
\end{eqnarray}
where $\langle \rangle_{\rm rad}$ denotes integration over radial variables only, and $\langle\Psi_{-}| H_{||}| \Psi_{-}\rangle=-\langle\Psi_{+}| H_{||} |\Psi_{+}\rangle$. The off-diagonal elements $\langle\Psi_{+}| H_{||}| \Psi_{-}\rangle$ and $\langle\Psi_{-}| H_{||} |\Psi_{+}\rangle$ vanish after integration over the angle $\phi$. By comparison with Eq.~(\ref{eq:gtensordef}), we get
\begin{equation}\label{gparallel}
g_{\parallel} = 8\pi \,c\, \left(\langle\psi_{1}| \rho |\psi_{4}\rangle_{\rm rad} - \langle\psi_{2}| \rho |\psi_{3}\rangle_{\rm rad} \right).
\end{equation}

%%%%%%%%%%%%%%%%%%%%%%%%%%%%%%%%%%%%%%%%%%%%%%%%%%%%%%%%%%%%%
\paragraph{Perpendicular field.}
%%%%%%%%%%%%%%%%%%%%%%%%%%%%%%%%%%%%%%%%%%%%%%%%%%%%%%%%%%%%%

In the perpendicular case, $H_{int}$ is given in Eq. \ref{Hperpx}. It is easy to verify that $\langle\Psi_{+}| H_{\perp} |\Psi_{+}\rangle = \langle\Psi_{+}| H_{\perp} |\Psi_{+}\rangle=0$. For the off-diagonal elements $\langle\Psi_{+}|H_{\perp}^{(1,2)}|\Psi_{-}\rangle=\langle\Psi_{-}|H_{\perp}^{(1,2)}|\Psi_{+}\rangle$, we obtain for the first interaction term (see Appendix \ref{apA2} for details)
\begin{eqnarray}
   \langle\Psi_{+}| H_{\perp}^{(1)}|\Psi_{-}\rangle
&=&-\frac{c}{2} B_{\rm x}  \langle\Psi_{+}| M_{\perp}(z)|\Psi_{-}\rangle\nonumber\\
&=& 2\pi \, c\, B_{\rm x} \langle\psi_{1}| z |\psi_{3}\rangle_{\rm rad} \,,
\label{gperp1}
\end{eqnarray}
and for the second interaction term 
\begin{eqnarray}
   \langle\Psi_{+}| H_{\perp}^{(2)}|\Psi_{-}\rangle
&=&-\frac{c}{2} B_{\rm x}  \langle\Psi_{+}| M_{\perp}(\rho,\sin(\phi))|\Psi_{-}\rangle \nonumber  \\
&=& \pi \, c \, B_{\rm x} \left(\langle\psi_{1}| \rho |\psi_{4} \rangle_{\rm rad} +  \langle\psi_{2}|\rho |\psi_{3} \rangle_{\rm rad} \right) \label{gperp2}
\end{eqnarray}
In total, by comparison with Eq.~(\ref{eq:gtensordef}) the magnetic moment in the perpendicular case yields
\begin{eqnarray}\label{gperp3}
    g_{\perp}&=& 4\pi \,c \, \left(\langle\psi_{1}| \rho |\psi_{4}\rangle_{\rm rad} +  \langle\psi_{2}| \rho |\psi_{3}\rangle_{\rm rad} +2  \langle\psi_{1}| z |\psi_{3}\rangle_{\rm rad} \right).
\end{eqnarray}
It is worth noting that in our numerical calculations we observe that the following equality always holds, within our numerical uncertainties:
\begin{equation}\label{gperp4}
   \langle\psi_{1}|  \rho |\psi_{4}\rangle_{\rm rad} + \langle\psi_{2}|  \rho |\psi_{3}\rangle_{\rm rad} = 2  \langle\psi_{1}|z |\psi_{3}\rangle_{\rm rad},
\end{equation}
so that the perpendicular $g$ tensor component could be simplified to
\begin{equation} \label{gperp-simple}
g_{\perp}= 16\pi \, c \, \langle\psi_{1}| z |\psi_{3}\rangle_{\rm rad}.    
\end{equation}
We also observed that in the atomic limit, where the $g$ factor becomes isotropic, the term $\langle\psi_{2}| \rho |\psi_{3}\rangle_{\rm rad}$ vanishes and we find
\begin{equation}\label{gatom}
     g_{||}^{\rm atom}= g_{\perp}^{\rm atom}
     %=g_{\rm xx}^{atom}=\mu_{\rm yy}^{atom} =\mu_{\rm zz}^{atom}   \\\nonumber
    = 8\pi \, c \, \langle\psi_{1}| \rho |\psi_{4}\rangle_{\rm rad}  = 16\pi \, c \,  \langle\psi_{1}| z |\psi_{3}\rangle_{\rm rad}.
\end{equation}
This identity, as well as the equality $\langle\psi_{2}| \rho |\psi_{3}\rangle_{\rm rad} = 0$, can be easily confirmed using the separation of angular and radial variables in the hydrogenic atom ground-state wavefunction (see e.g.~\cite{Bethe1957}).

%%%%%%%%%%%%%%%%%%%%%%%%%%%%%%%%%%%%%%%%%%%%%%%%%%%%%%%
\subsection{Numerical minmax FEM Method}\label{FEMM}
%%%%%%%%%%%%%%%%%%%%%%%%%%%%%%%%%%%%%%%%%%%%%%%%%%%%%%%

The evaluation of relativistic $g$ tensor components  is performed using high-precision numerical wave functions obtained from resolution of the Dirac equation in a two-center potential by the FEM \cite{Kullie20041,Kullie2022}. This method has allowed calculation of the relativistic ground-state energy of H$_2^+$ with about 10$^{-23}$ accuracy~\cite{Kullie2024}. These results have been confirmed by the excellent agreement with similarly accurate results obtained by other methods~\cite{Nogueira2022,Nogueira2023}.   
 
We briefly recall the main features of our approach. It makes use of the minmax principle~\cite{Dolbeault20002}, which is based on the elimination of the small component from the Dirac equation, leading to a nonlinear eigenvalue problem that is solved iteratively. The nonlinearity is (even for heavy systems) weak, and does not cause any problem in iterative linearized computations of the eigenvalues. Furthermore, in the nonrelativistic limit $c\to \infty$ the resulting equation transforms directly into the Schr\"odinger one. The point of the minmax approach is that the computed eigenvalues follow a minimization principle; all levels of the computed spectrum approach the exact Dirac eigenvalue from above as the discretization of space is refined by increasing the number of elements in the grid. Moreover, the spectrum is free from spurious states.

In view of the axial symmetry around the internuclear axis, the angular coordinate $\phi$ is separated, as shown in Eqs.~(\ref{eq:sep-angle}), (\ref{eq:timerevWF}) and the equation verified by the radial wavefunction is written in the prolate spheroidal coordinate system. In order to deal with the singularity of the wavefunction in the vicinity of the point nuclei, a further singular coordinate transformation is performed, characterized by an integer order $\nu$~\cite{Kullie2001}. The higher $\nu$, the denser the grid near the nuclei to ensure a better representation of the singularity. According to the FEM, each component of the wavefunction is expanded as a sum over the triangular grid elements; the shape functions defined in each elements are 2-variable complete polynomials of order $p$. The size of the grid is characterized by $D_{\rm max}$, defined as the distance between one of the nuclei to a point on the outermost ellipse, where this distance is perpendicular to the internuclear axis.

In this work, we use similar parameter values as in~\cite{Kullie2022}.  For the regularization of the singularity near the nuclei, we choose $\nu=4$ or $6$, which is sufficient for the desired accuracy. Regarding the polynomial order of the shape functions, we exclusively use $p=10$. The extension of the domain is $D_{\rm max}=50$ a.u. (unless otherwise stated). The parameters used in the calculations are given in the table captions. For more details, we refer to our previous work \cite{Kullie2022,Kullie2024}.

In our FEM approach, we perform the computation for a series of successive grids in order to study the convergence of the results and estimate uncertainties. The solution obtained with the finest grid is used to calculate the final values of the $g$ tensor components.

The accuracy achieved in the present work on the energy levels is on the same order as in Ref. \cite{Kullie2022}. The total energies have estimated fractional uncertainties of a few times $10^{-20}$, whereas the fractional uncertainty of the purely relativistic shift is on the order of $10^{-16}$.

In all the numerical calculations, except otherwise noted, we use the CODATA-2018 value $c = 1/\alpha = 137.035999084$. Note that using the latest (CODATA-2022) value would shift the results by only a few parts in $10^{-14}$, which is well below the estimated order of magnitude of the largest yet uncalculated QED contributions to the $g$ factor.

%%%%%%%%%%%%%%%%%%%%%%%%%%%%%%%%%%%%%%%%%%%%%%%%%%%%%%%%%%%%%%%%%
\subsection{Atomic limit}\label{sec:result}
%%%%%%%%%%%%%%%%%%%%%%%%%%%%%%%%%%%%%%%%%%%%%%%%%%%%%%%%%%%%%%%%% 
As a first test for our calculations, we show in table \ref{tab:g_Hatom} the calculated values of $g_{||}=g_{\perp}$ for the H atom, which can be easily represented in our approach by setting a zero value for the charge of the second center ($Z_2=0$), and compare it with the exact result~\cite{Breit1928}: $g_{\rm _{Breit}}=(2/3)(1+2\sqrt{1-Z^2/c^2})=1.9999644986243560008411175$.

Table \ref{tab:g_Hatom} shows that our method can achieve a highly accurate result.
%%%%%%%%%%%%%%%%%%%%%%%%%%%%%%%%%%%%%%%%%%%%%%%%%%%%%%%%%%%%%%%%
\begin{table}[htpb]
\begin{tabular*}{\linewidth}{@{\extracolsep{\fill}} llll}
\hline\hline \\ [-2.5ex]
$N_g$ & \hspace{1.0cm} $g_{||}$ [Eq.~(\ref{gparallel})] & \hspace{1.0cm}  $g_{\perp}$ [Eqs.~(\ref{gperp3},\ref{gperp-simple})] & \hspace{1.0cm} $g_s$ [Eq. \ref{eq:link gs}]
\\ \hline \\ [-2.0ex]
  441  & 1.9999644991701934676777621 &  1.9999641624258261032669356 & 1.9999642746739485580705444\\
 6561  & 1.9999644986243560008454195 &  1.9999644986243560007464186 & 1.9999644986243560007794189\\
 10201 & 1.9999644986243560008417426 &  1.9999644986243560008415023 & 1.9999644986243560008415824\\
 19881 & 1.9999644986243560008411175 &  1.9999644986243560008411170 & 1.9999644986243560008411172\\
\hline \hline
\end{tabular*}
\caption{Calculated values of the H atom relativistic $g$ factor as a function of the grid point number $N_g$. 
The exact value ($g_{\rm Breit}$)~\cite{Breit1928} is given in the text). The error of the calculated value in the last line is about $3 \times 10^{-25}$. FEM parameters: $p=10$, $\nu=4$ (see text in Sec.~\ref{FEMM} for details).}    
\label{tab:g_Hatom}
\end{table}
%%%%%%%%%%%%%%%%%%%%%%%%%%%%%%%%%%%%%%%%%%%%%%%%%%%%%%%%%%%%

A further test for our calculation is to approach the H atom limit by considering the ${\rm H}_{2}^{+}$ molecular in the limit of large internuclear distances $R$. This is described in the Appendix \ref{app-largeR}.
%%%%%%%%%%%%%%%%%%%%%%%%%%%%%%%%%%%%%%%%%%%%%%%%%%%%%%%%BIB
\begin{table}[htpb]
  \begin{tabular*}{\linewidth}{@{\extracolsep{\fill}} lll}
\hline\hline\\ [-2.0ex] 
$N_g$ &  $(1-g_{\parallel}/g_{\rm e}) \times 10^6$&  $(1-g_{\perp}/g_{\rm e}) \times 10^6$
\\ \hline\\ [-1.5ex] 
6561  & {\bf 19.7704625310577}6811 & {\bf 21.0609403518}40950
\\
10201 & {\bf 19.77046253105778}892 & {\bf 21.06094035185}0300
\\
14641 & {\bf 19.770462531057786}13 & {\bf 21.06094035185}2908
\\
19881 & {\bf 19.7704625310577860}6 & {\bf 21.06094035185}3773
\\
25921 & {\bf 19.7704625310577860}4 & {\bf 21.060940351854}098 
\\
32761 & {\bf 19.77046253105778604} & {\bf 21.060940351854}232
\\
32761 (with $\nu = 6$) &{\bf 19.77046253105778604} & {\bf 21.06094035185435}0
\\ \hline
32761 (with $c=137.03602$) & {19.7704564958733512} & {21.060933922739}98
\\
$\alpha^2$ correction (Ref.\,\cite{Hegstrom1979}) & 19.7705 & 21.0610
\\
$\alpha^2$ correction (this work) & 19.77046297 & 21.06095689 
\\\hline \hline
\end{tabular*}
\caption{\label{tab:g_H2+_1} 
$\Delta g\,(=1-g/g_{\rm e}$) values (in units of $10^{-6}$) for the H$_2^+$ electronic ground state at $R = 2$~a.u. The approximation $g_{\rm e} = 2$ is used in the calculation of $\Delta g$. Upper part: values as a function of the grid size number $N_g$ using the CODATA-2018 value of $c = 1/\alpha$, with FEM parameters $\nu=4$ and $p=10$. The last value is calculated with $\nu=6$. Bold digits are converged. Lower part: comparison with calculations of the $\alpha^2$-order relativistic correction; in this part, all calculations are performed with $c=137.036020$.}
\end{table}

%%%%%%%%%%%%%%%%%%%%%%%%%%%%%%%%%%%%%%%%%%%%%%%%%%%%%%%%%%%%%
\subsection{\texorpdfstring{${\rm H}_{2}^{+}$} \textsc{} molecular ion: convergence study}
%%%%%%%%%%%%%%%%%%%%%%%%%%%%%%%%%%%%%%%%%%%%%%%%%%%%%%%%%%%%%

We now present calculations of the $g$ tensor of the ${\rm H}_{2}^{+}$ electronic ground state at the internuclear distance $R=2$~a.u. and compare them with the available results from the literature~\cite{Hegstrom1979}, which only include the leading-order ($\alpha^2$) relativistic correction (see Sec.~\ref{sec-gHegstrom}, Eqs.~(\ref{eq:g2perp-nonrec}-\ref{eq:g2par-nonrec})). Our results are shown in Table~\ref{tab:g_H2+_1}; in order to ease comparison with Ref.~\cite{Hegstrom1979}, they are given in terms of the relative deviation with respect to the free-electron $g$ factor (under the approximation $g_{\rm e} = 2$).

The first part of the table gives our results for a sequence of grid point numbers $N_g$, illustrating their convergence. The achieved absolute accuracy using the a singular coordinate transformation of order $\nu=4$ is better than $10^{-23}$ for $g_{\parallel}$ and $10^{-20}$ for $g_{\perp}$. Using $\nu=6$ allows further improving the accuracy of $g_{\perp}$.

In the second part of the table, we compare our results (using the largest grid) with those of Ref.~\cite{Hegstrom1979}, and with our own (more precise) calculations of the $\alpha^2$-order correction (see Sec~\ref{sec:ad-calc} for details). For this, we redid the calculations using the same value of $c$ as in~\cite{Hegstrom1979} ($c=137.03602$). Good agreement is observed at the $10^{-10}$ precision level of the results of~\cite{Hegstrom1979}. Our calculations of the $\alpha^2$-order correction allow for a more precise comparison. The relative difference $(g_{\rm Dirac} - g_{\alpha^2})/g_{\rm e}$ amounts to $-6.5 \times 10^{-12}$ ($-2.3 \times 10^{-11}$) for the parallel (perpendicular) component. We note that these values are markedly smaller than the $\alpha^4$-order correction to the hydrogen atom $g$ factor obtained by expanding the Breit formula, $\Delta g^{(4)}/g_{\rm e} = -(Z\alpha)^4/12 = -2.4 \times 10^{-10} Z^4$. We further checked that in the atomic limit, $(g_{\rm Dirac} - g_{\alpha^2})/g_{\rm e}$ is as expected very close to the above value, with $Z=2$ ($Z=1$) in the small-$R$ (large-$R$) limit.

In the remainder of the present work, unless otherwise specified, calculations of the relativistic $g$ factor are carried out with the same accuracy as shown in Table~\ref{tab:g_H2+_1}, using the finest grid ($N_g=32761$) and the singular coordinates order $\nu=6$ (with $p=10$, $D_{\rm max}=50$~a.u.).

%%%%%%%%%%%%%%%%%%%%%%%%%%%%%%%%%%%%%%%%%%%%%%%%%%%%%%%%%%%%%
\subsection{Relativistic \texorpdfstring{$g$}\textsc{} factor of \texorpdfstring{H$_2^+$}\textsc{} and  \texorpdfstring{HD$^+$}\textsc{} rovibrational states}
%%%%%%%%%%%%%%%%%%%%%%%%%%%%%%%%%%%%%%%%%%%%%%%%%%%%%%%%%%%%%

In the adiabatic framework, after performing calculations for a range of values of the internuclear distance $R$, the $g$ tensor components of a given RV state $(\nu,N)$ are calculated by averaging over the RV wavefunction $f_{\nu,N} (R)$: 
\begin{equation}\label{intRV}
g_{\parallel(\perp)} (\nu,N)=\int_{0}^{R_{\infty}} dR\, g_{\parallel(\perp)}(R)\, [f_{\nu,N}(R)]^2 \,,
\end{equation}
where $R_{\infty}$ is chosen sufficiently large for the $R>R_{\infty}$ remainder to be negligibly small. In practice, we use $R_{\infty} = 20$~a.u., which is sufficient for the range of RV states considered here.

At this stage of development of the $g$-factor theory, the RV degree of freedom can be described in a nonrelativistic way, because leading relativistic corrections to the nuclear motion are of order $\alpha^2 (m_{\rm e}/M)$. Since leading corrections to the $g$ factor are of order $\alpha^2$, relativistic nuclear effects would produce corrections of leading order $\alpha^4 (m_{\rm e}/M) \sim 10^{-12}$, which is beyond our current precision goal.

However, the indirect effect of relativistic corrections to the electronic energy curve on the RV wavefunction is of order $\alpha^2$, leading to an $\alpha^4$ correction to the $g$ factor. A similar ``vibrational'' term occurs in the relativistic correction to the RV energy levels~\cite{Korobov2017}. In order to calculate correctly the $\alpha^4$-order relativistic correction, it is thus mandatory to include leading-order relativistic corrections in the electronic energy curve that is used to solve the nuclear Schr\"odinger equation. Using perturbation theory, one could isolate the $\alpha^4$-order contribution, similarly to what was done in~\cite{Korobov2017}. In the present work, we instead perform an all-order calculation, i.e. we solve the nuclear Schr\"odinger equation using the Dirac electronic energy curve, including the non-relativistic adiabatic correction~\cite{Wolniewicz1980,Carrington1989}. Relativistic corrections to the adiabatic correction are of order $\alpha^2 (m_{\rm e}/M)$, producing corrections of leading order $\alpha^4 (m_{\rm e}/M)$ to the $g$ factor, and can thus be neglected here.

Summarizing, the RV wavefunction is obtained by solving the radial equation
\begin{equation}
\left( -\frac{1}{2\mu_N} \frac{d^2}{dR^2} + U(R) - E_{\nu,N} + \frac{N(N+1)}{R^2} \right) f_{\nu,N}(R) = 0 \,,
\end{equation}
where $\mu_N = M_1 M_2/(M_1 + M_2)$ is the reduced mass of the nuclei, and
\begin{equation}
U(R) = E_{\rm Dirac}(R) + \frac{Z_1 Z_2}{R} + E_{\rm ad}(R). \label{eq:pot-curve}
\end{equation}
For the Dirac energies $E_{\rm Dirac}(R)$ we use the results of~\cite{Kullie2022,Kullie2024}. The expression of the adiabatic correction $E_{\rm ad}(R)$ can be found in~\cite{Wolniewicz1980,Carrington1989}.

%%%%%%%%%%%%%%%%%%%%%%%%%%%%%%%%%%%%%%%%%%%%%%%%%%%%%%%%
\subsubsection{Effect of relativistic corrections to RV wavefunctions}\label{sec:rvs}
%%%%%%%%%%%%%%%%%%%%%%%%%%%%%%%%%%%%%%%%%%%%%%%%%%%%%%%%

In order to illustrate the effect of including relativistic corrections to the electronic energy curve for the computation of RV wavefunctions, we now compare the scalar $g$ factor values obtained using ``relativistically corrected'' (denoted by $f^{\rm rel}_{\nu,N}$) and purely nonrelativistic (denoted by $f^{\rm nrel}_{\nu,N}$) RV wavefunctions in Eq.~(\ref{intRV}). The latter are obtained by replacing the Dirac energy curve $E_{\rm Dirac}(R)$ with the Schr\"odinger one, $E_{\rm S}(R)$, in Eq.~(\ref{eq:pot-curve}). For this, the two-center Schr\"odigner equation is solved with 20-digit accuracy using the variational method described in Sec.~\ref{sec:ad-calc}. For both the Dirac and Schr\"odinger cases, the numerical integration over $R$ is done with a step size of $0.05$ a.u.

\begin{figure}[htpb]
\includegraphics[width=6.0cm]{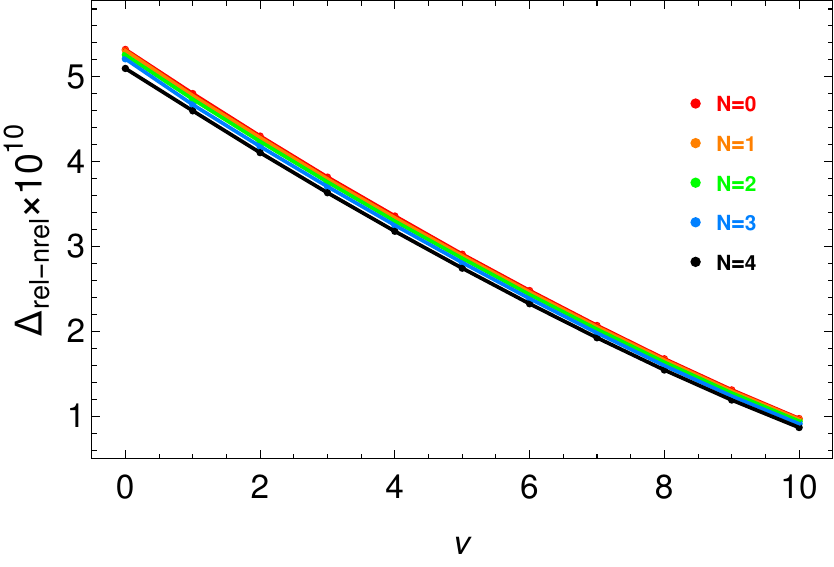}
\hspace{1cm}
 \vspace{-10pt}
 \caption{\label{fig:diffnu0} 
 Variation of the relativistic shift $\Delta_{\rm rel-nrel}$ of the scalar $g$ factor (see Table~\ref{tab:Lnu00}) as a function of the vibrational state $\nu$, for $N=0-4$.}
 \end{figure}

The difference $\Delta_{\rm rel-nrel}$ between the relativistic and nonrelativistic values is
\begin{equation}
\Delta_{\rm rel-nrel} = \int_{0}^{R_{\infty}} dR\, g_s(R)\, \left[ (f^{\rm rel}_{\nu,N}(R))^2 - (f^{\rm nrel}_{\nu,N}(R))^2 \right]. \label{eq:deltarel}
\end{equation}
The values of $\Delta_{\rm rel-nrel}$ for the first few RV states of H$_2^+$ are given in Table~\ref{tab:Lnu00} (in units of $10^{-10}$); they are compatible with the order $\alpha^4 \sim 2.8 \times 10^{-9}$. Furthermore, Fig.~\ref{fig:diffnu0} shows the dependence of $\Delta_{\rm rel-nrel}$ on $\nu$ for $N=0,4$. The relativistic effect has a trend towards a lower value with increasing $\nu$, and only weakly depends on $N$.

In the following, we use the relativistically corrected RV wavefunctions $f^{\rm rel}_{\nu,N}(R)$ in our calculations of the Dirac $g$ factor.
%%%%%%%%%%%%%%%%%%%%%%%%%%%%%%%
\begin{table}[htpb]
    \begin{tabular}{ccc}
        %\begin{array}{ccccccc}
        \hline\hline \\ [-2.0ex]
      & $\nu=0$      & $\nu=1$ \\
      \hline\\ [-2.0ex]
$N=0$ & -5.33  & -4.80 \\
$N=1$ & -5.30  & -4.78 \\
$N=2$ & -5.26  & -4.74 \\
$N=3$ & -5.19  & -4.67 \\
$N=4$ & -5.10  & -4.59 \\
\hline\hline
    \end{tabular}
\caption{Relativistic effect $\Delta_{\rm rel-nrel}$ [see Eq.~(\ref{eq:deltarel})], in units of $10^{-10}$, for the RV states $N=0-4, \nu=0,1 $ of H$_2^+$. } \label{tab:Lnu00}
\end{table}
%%%%%%%%%%%%%%%%%%%%%%%%%%%%%%%%%%
%%%%%%%%%%%%%%%%%%%%%%%%%%%%%%%%%%

%%%%%%%%%%%%%%%%%%%%%%%%%%%%%%%%%%%%%%%%%%%%%%%%%%%%%%%%
\subsubsection{Comparison with literature values}
%%%%%%%%%%%%%%%%%%%%%%%%%%%%%%%%%%%%%%%%%%%%%%%%%%%%%%%%

As a first test, we compare our values of the relativistic correction $g$ tensor components with previous calculations of the $\alpha^2$ order correction for the RV states ($\nu=0,N=0,1$) of ${\rm H}_{2}^{+}$, see Table~\ref{tab:Tcomp1}. Similarly to Table~\ref{tab:g_H2+_1}), results are given in terms of relative deviations with respect to the free-electron $g$ factor in order to allow direct comparison with Refs.~\cite{Hegstrom1979,Karr2021}. The values of ``this work'' are obtained using the CODATA-2018 value of $c=\alpha^{-1}$ (also used in~\cite{Karr2021}), although a different value was used in~\cite{Hegstrom1979}. Indeed, the shift of the $g$ factor value induced by the change of $c$ is in the 10$^{-12}$ range (as can be seen in Table~\ref{tab:g_H2+_1}), which is well below the precision of the comparison.

Our Dirac results are in good agreement with the leading-order calculation of Ref.~\cite{Hegstrom1979}. The slight deviation of about $10^{-9}$ is not due to higher-order relativistic corrections, but comes from the precision limit of Hegstrom's results, as shown by our more precise calculations (described in Sec.~\ref{sec:ad-calc}). The difference between our Dirac and $\alpha^2$ results amounts to about $2.6 \times 10^{-10}$ and corresponds to the $\alpha^4$-order (and higher) relativistic correction, $\Delta g_{\rm nonrec}^{\rm rel(4+)}$ (see Eqs.~(\ref{eq:grel4perp}-\ref{eq:grel4par})). The comparison for the $(\nu=0,N=1)$ state shows that higher-order relativistic corrections to the tensor component $g_t$ are smaller than $10^{-11}$.

The three-body result of Ref.~\cite{Karr2021} includes non-adiabatic and recoil corrections of order $\alpha^2$, which explains the observed deviations. The differences of about $5 \times 10^{-9}$ for $g_s$ and $3 \times 10^{-9}$ for $g_t$ correspond to the terms $\Delta g_{\rm s-3body}^{(2)}$, $\Delta g_{\rm t-3body}^{(2)}$ defined in Eqs.~(\ref{eq:gs3b}-\ref{eq:gt3b}), see Sec.~\ref{sec:3body corr} for more details.

%%%%%%%%%%%%%%%%%%%%%%%%%%%%%%%%%%%%%%%%%%%%%%%%%%%%%%%%
\begin{table}[htpb]
  \begin{tabular*}{\linewidth}{@{\extracolsep{\fill}} lllll} \hline\hline \\ [-2.0ex]
  & $(1\!-\!g_{\parallel}/g_{\rm e}) \times 10^6$ & $(1\!-\!g_{\perp}/g_{\rm e}) \times 10^6$ & $(1\!-\!g_s/g_{\rm e}) \times 10^6$ & $(1\!-\!g_t/g_{\rm e}) \times 10^6$
  \\ \hline  \\ [-2.0ex]
%%%%%%%%%%%%%%%  
\multicolumn{5}{c}{$\nu=0, N=0$}\\\hline \\ [-2.0ex]
Dirac (this work)                          & 19.52743 & 20.77697 & 20.36046   & -\\
$\alpha^2$ (nonrecoil) (this work)         & 19.52718 & 20.77671 & 20.36020   & -\\
$\alpha^2$ (nonrecoil) \cite{Hegstrom1979} & 19.526   & 20.776   & 20.359     & -\\
$\alpha^2$ (with recoil) \cite{Karr2021}   & -        & -        & 20.3552762 & -
\\\hline \\ [-2.0ex]
%%%%%%%%%%%%%%%
\multicolumn{5}{c}{$\nu=0, N=1$}\\\hline \\ [-2.0ex]
Dirac (this work)   & 19.50927    & 20.75706   & 20.34113 & -0.52611    \\
$\alpha^2$ (nonrecoil) (this work) & 19.50902  & 20.75680 & 20.34088 & -0.52611 \\
$\alpha^2$ (nonrecoil) \cite{Hegstrom1979} & 19.508      & 20.756     & 20.340     & -0.526      \\
$\alpha^2$ (with recoil) \cite{Karr2021}     & 19.5000329  & 20.7539086 & 20.3359500 & -0.5286804
 \\\hline\hline
\end{tabular*}
\caption{\label{tab:Tcomp1} 
$\Delta g\,(=1-g/g_{\rm e}$) values (in units of $10^{-6}$) for the $(\nu=0,N=0,1)$ RV states of H$_2^+$. The approximation $g_{\rm e} = 2$ is used in the calculation of $\Delta g$.}
\end{table}
%%%%%%%%%%%%%%%%%%%%%%%%%%%%%%%%%%%%%%%%%%%%%%%%%%%%%%%%%%

%%%%%%%%%%%%%%%%%%%%%%%%%%%%%%%%%%%%%%%%%%%%%%%%%%%%%%%%%%
\subsubsection{Relativistic  \texorpdfstring{$g$}\textsc{} factor values}
%%%%%%%%%%%%%%%%%%%%%%%%%%%%%%%%%%%%%%%%%%%%%%%%%%%%%%%%%%

We give here our results the relativistic $g$ tensor components for the first vibrational states ($\nu = 0-4$) in H$_2^+$ (Table~\ref{tab:grel-H2+}) and HD$^+$ (Table~\ref{tab:grel-HD+}). Their precision is limited by the numerical integration over the internuclear distance $R$, all the more as the vibrational level is increased. The uncertainty rises from $7-9 \times 10^{-12}$ for $\nu=0$ to $5-7 \times 10^{-10}$ for $\nu=4$. Beyond that, the precision is too low to be sensitive to the higher-order relativistic correction, which is why we do not report results for higher vibrational states here.

While the precision could be improved by using a smaller step size for the integration, it is important to stress that this precision limit does not affect the final $g$-factor predictions presented in the next section. Indeed, the relativistic $g$ factor is only used to extract the higher-order relativistic correction by subtracting the leading-order terms, see Eqs.~(\ref{eq:grel4perp}-\ref{eq:grel4par}). Then, the relative inaccuracy of the numerical integration only affects the small higher-order remainder, leading to a very small absolute uncertainty.

%%%%%%%%%%%%%%%%%%%%%%%%%%%%%%%%%%%%%%%%%%%%%%%%%%%%%%%%%%
\begin{table}[htpb]
\begin{tabular}{clllll}
\hline \hline \\ [-2.0ex]
 \text{} & \text{N=0} & \text{N=1} & \text{N=2} & \text{N=3} & \text{N=4} \\
\hline \\ [-2.0ex]
 \multirow{2}{*}{\text{$\nu $=0}} & 1.999\,960\,945\,135(7) & 1.999\,960\,981\,451(7) & 1.999\,961\,053\,631(7) & 1.999\,961\,160\,783(7) & 1.999\,961\,301\,611(7) \\
& 1.999\,958\,446\,066(8) & 1.999\,958\,485\,880(8) & 1.999\,958\,565\,041(8) & 1.999\,958\,682\,631(8) & 1.999\,958\,837\,309(8) \\
\hline \\ [-2.0ex]
 \multirow{2}{*}{\text{$\nu $=1}} & 1.999\,961\,910\,57(4) & 1.999\,961\,944\,31(4) & 1.999\,962\,011\,34(4) & 1.999\,962\,110\,83(4) & 1.999\,962\,241\,54(4) \\
& 1.999\,959\,578\,72(5) & 1.999\,959\,615\,93(5) & 1.999\,959\,689\,90(5) & 1.999\,959\,799\,75(5) & 1.999\,959\,944\,22(5) \\
\hline \\ [-2.0ex]
 \multirow{2}{*}{\text{$\nu $=2}} & 1.999\,962\,790\,0(1) & 1.999\,962\,821\,3(1) & 1.999\,962\,883\,3(1) & 1.999\,962\,975\,4(1) & 1.999\,963\,096\,4(1) \\
 & 1.999\,960\,622\,3(2) & 1.999\,960\,657\,0(2) & 1.999\,960\,726\,0(2) & 1.999\,960\,828\,3(2) & 1.999\,960\,962\,9(1) \\
\hline \\ [-2.0ex] 
\multirow{2}{*}{\text{$\nu $=3}} & 1.999\,963\,586\,3(3) & 1.999\,963\,615\,1(3) & 1.999\,963\,672\,3(3) & 1.999\,963\,757\,2(3) & 1.999\,963\,868\,6(3) \\
 & 1.999\,961\,579\,4(3) & 1.999\,961\,611\,7(3) & 1.999\,961\,675\,7(3) & 1.999\,961\,770\,8(3) & 1.999\,961\,895\,7(3) \\
\hline \\ [-2.0ex] 
\multirow{2}{*}{\text{$\nu $=4}} & 1.999\,964\,301\,2(5) & 1.999\,964\,327\,7(5) & 1.999\,964\,380\,2(5) & 1.999\,964\,458\,0(5) & 1.999\,964\,560\,1(5) \\
& 1.999\,962\,451\,8(6) & 1.999\,962\,481\,6(6) & 1.999\,962\,540\,9(6) & 1.999\,962\,628\,8(6) & 1.999\,962\,744\,2(6) 
\\
\hline \hline \\ [-2.0ex]
\end{tabular}
\caption{Values of the components of the relativistic (Dirac) $g$ tensor for RV states of H$_2^+$. The first and second lines correspond to the parallel ($\Delta g_{\parallel-{\rm nonrec}}^{\rm rel}$) and perpendicular ($\Delta g_{\perp-{\rm nonrec}}^{\rm rel}$) components.} 
\label{tab:grel-H2+}
\end{table}
%%%%%%%%%%%%%%%%%%%%%%%%%%%%%%%%%%%%%%%%%%%%%%%%%%%%%%%%%%
\begin{table}[htpb]
\begin{tabular}{clllll}
\hline \hline \\ [-2.0ex]
 \text{} & \text{N=0} & \text{N=1} & \text{N=2} & \text{N=3} & \text{N=4} \\
\hline \\ [-2.0ex]
 \multirow{2}{*}{\text{$\nu $=0}} & 1.999\,960\,874\,868(8) & 1.999\,960\,902\,263(8) & 1.999\,960\,956\,797(8) & 1.999\,961,037\,963(8) & 1.999\,961\,145\,020(8) \\
& 1.999\,958\,364\,268(9) & 1.999\,958\,394\,288(9) & 1.999\,958\,454\,065(9) & 1.999\,958\,543\,076(9) & 1.999\,958\,660\,554(9) \\
\hline \\ [-2.0ex]
 \multirow{2}{*}{\text{$\nu $=1}} & 1.999\,961\,721\,42(5) & 1.999\,961\,747\,12(5) & 1.999\,961\,798\,29(5) & 1.999\,961\,874\,44(5) & 1.999\,961\,974\,85(5) \\
& 1.999\,959\,356\,00(6) & 1.999\,959\,384\,32(6) & 1.999\,959\,440\,70(6) & 1.999\,959\,524\,65(6) & 1.999\,959\,635\,42(6) \\
\hline \\ [-2.0ex]
 \multirow{2}{*}{\text{$\nu $=2}} & 1.999\,962\,503\,0(2) & 1.999\,962\,527\,1(2) & 1.999\,962\,575\,0(2) & 1.999\,962\,646\,2(2) & 1.999\,962\,740\,2(2) \\
 & 1.999\,960\,280\,5(2) & 1.999\,960\,307\,1(2) & 1.999\,960\,360\,2(2) & 1.999\,960\,439\,2(2) & 1.999\,960\,543\,5(2) \\
\hline \\ [-2.0ex] 
\multirow{2}{*}{\text{$\nu $=3}} & 1.999\,963\,221\,5(4) & 1.999\,963\,243\,9(4) & 1.999\,963\,288\,7(3) & 1.999\,963\,355\,2(3) & 1.999\,963\,442\,9(3) \\
 & 1.999\,961\,139\,4(4) & 1.999\,961\,164\,5(4) & 1.999\,961\,214\,3(4) & 1.999\,961\,288\,6(4) & 1.999\,961\,386\,5(4) \\
\hline \\ [-2.0ex] 
\multirow{2}{*}{\text{$\nu $=4}} & 1.999\,963\,878\,3(6) & 1.999\,963\,899\,2(6) & 1.999\,963\,940\,9(6) & 1.999\,964\,002\,8(6) & 1.999\,964\,084\,3(6) \\
& 1.999\,961\,934\,2(7) & 1.999\,961\,957\,7(7) & 1.999\,962\,004\,4(7) & 1.999\,962\,073\,9(7) & 1.999\,962\,165\,6(7) \\
\hline \hline \\ [-2.0ex]
\end{tabular}
\caption{Same as Table~\ref{tab:grel-H2+}, for RV states of HD$^+$.} \label{tab:grel-HD+}
\end{table}
%%%%%%%%%%%%%%%%%%%%%%%%%%%%%%%%%%%%%%%%%%%%%%%%%%%%%%%%%%

%%%%%%%%%%%%%%%%%%%%%%%%%%%%%%%%%%%%%%%%%%%%%%%%%%%%%%%%
\section{Bound-electron  \texorpdfstring{$g$}\textsc{} factor: numerical results} \label{sec:results}
%%%%%%%%%%%%%%%%%%%%%%%%%%%%%%%%%%%%%%%%%%%%%%%%%%%%%%%%%%

In this section, we briefly describe the numerical calculation of all the contributions to the $g$ tensor (see Sec.~\ref{sec:corrections}), apart from the Dirac contribution, which was discussed in Sec.~\ref{sec:grel}. Then, we present our complete results for a range of RV states of H$_2^+$ and HD$^+$ in Tables~\ref{tab:final-g-H2+} and \ref{tab:final-g-HD+}.

%%%%%%%%%%%%%%%%%%%%%%%%%%%%%%%%%%%%%%
\subsection{Three-body calculations}
%%%%%%%%%%%%%%%%%%%%%%%%%%%%%%%%%%%%%%

The complete $(Z\alpha)^2$-order relativistic correction including recoil terms (see Sec.~\ref{sec:g2tot}, Eqs.~(\ref{eq:gs2tot}-\ref{eq:gt2tot})) is calculated in a three-body formalism following the approach of~\cite{Karr2021}. It is based on high-precision resolution of the three-body Schr\"odinger equation using a variational expansion in a basis set of exponential functions of interparticle distances with pseudo-randomly chosen exponents~\cite{Korobov2000}. The only difference between the results shown here and those of~\cite{Karr2021} is that we take into account the exact dependence of the correction terms on the free-electron $g$ factor $g_{\mathrm{e}}$, whereas the approximation $g_{\mathrm{e}}=2$ had previously been used. We refer to~\cite{Karr2021} for a description of the numerical method and implementation details.

Our results are given in columns 3 and 8 of Tables~\ref{tab:final-g-H2+} and \ref{tab:final-g-HD+}. The same accuracy level as in~\cite{Karr2021} is achieved: the absolute uncertainty of $\Delta g_s^{(2)}$ and $\Delta g_t^{(2)}$ is smaller than $10^{-13}$, meaning that all the digits shown in the Tables are converged.

%%%%%%%%%%%%%%%%%%%%%%%%%%%%%%%%%%%%%%
\subsection{Adiabatic calculations} \label{sec:ad-calc}
%%%%%%%%%%%%%%%%%%%%%%%%%%%%%%%%%%%%%%

The adiabatic approximation is used to calculate nonrecoil corrections of order $\alpha^2$ (see Sec.~\ref{sec-gHegstrom}, Eqs.~(\ref{eq:g2perp-nonrec}-\ref{eq:g2par-nonrec})), $\alpha^3$ (Sec.~\ref{sec:g3}) and $\alpha^5$ (Sec.~\ref{sec:g5}). We recall that the adiabatic $\alpha^2$ correction is not directly used in the $g$ factor calculations (for which we instead use the complete three-body calculation described in the previous section), but has to be calculated to allow extracting the higher-order relativistic correction from the Dirac result, as explained in Sec.~\ref{sec:grelho}.

The electronic two-center Schr\"odinger equation is solved using the method introduced in~\cite{Tsogbayar:2006}. The $1s\sigma_g$ electronic wavefunction is expanded as follows:
\begin{equation}
\psi(\mathbf{r}) = \sum_{i=1}^{N_b} C_i \, \left( e^{-\alpha_i r_1 - \beta_i r_2} + e^{-\beta_i r_1 - \alpha_i r_2} \right)\,
\end{equation}
where the real exponents $\alpha_i$ and $\beta_i$ are generated in a pseudorandom manner in several intervals (typically 2 or 3). The calculation of matrix elements can be reduced to evaluation of integrals of the type
\begin{equation}
\Gamma_{lm} (\alpha,\beta) = \int r_1^{l-1} r_2^{m-1} e^{-\alpha r_1 - \beta r_2} \, d^3r.
\end{equation}
In the matrix elements of the Schr\"odinger Hamiltonian, the integers $l$, $m$ are non-negative. $\Gamma_{lm}$ can then be calculated from $\Gamma_{00}$ using a recurrence relation~\cite{Tsogbayar:2006}. However, for some operators (for example $p_z^2$, which appears in Eqs.~(\ref{eq:g3perp}-\ref{eq:g3par})), the matrix elements are singular in the limit $r_1,r_2 \to 0$ and need to be regularized. This is done by introducing a finite cutoff $r_0$ in the integrals and cancelling explicitly the divergent terms.

All the operator expectation values are obtained with at least 9-digit accuracy, using basis sizes $N_b$ between 100 and 150. The numerically most difficult term is the second-order term of Eq.~(\ref{eq:dg2perpC}), for which a typical basis size $N'_b \sim 250$ is used for representation of intermediate states. This high accuracy is only required for the $\alpha^2$-order contributions, in order to extract the higher-order remainder after subtraction from the Dirac result.

Results are shown in columns 4-6 and 9-10 of Tables~\ref{tab:final-g-H2+} and \ref{tab:final-g-HD+}, and the vibrational state dependence of the higher-order relativistic correction $\Delta g_{s-{\rm nonrec}}^{{\rm rel}(4+)}$ is shown in Fig.~\ref{fig:diffnu1} (a).

%%%%%%%%%%%%%%%%%%%%%%%%%%%%%%%%%%%%%%
\subsection{Three-body correction of order  \texorpdfstring{$\alpha^2$} \textsc{}} \label{sec:3body corr}
%%%%%%%%%%%%%%%%%%%%%%%%%%%%%%%%%%%%%%

\begin{figure}[htpb]
(a)
\includegraphics[width=6.0cm]{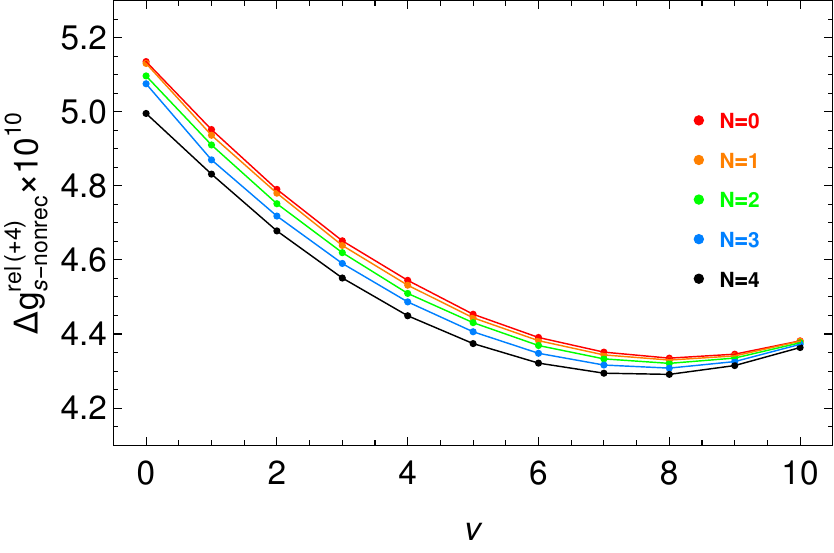}
\hspace{1cm}
(b)
\includegraphics[width=6.0cm]{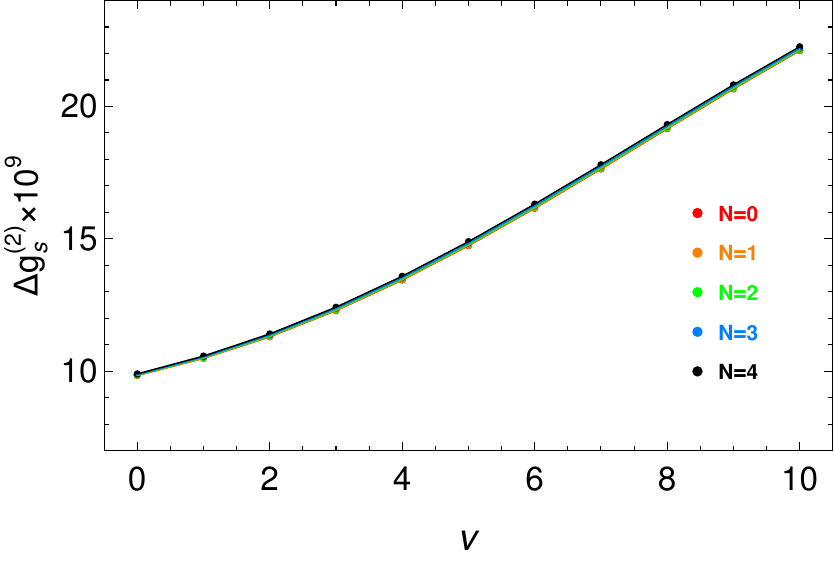}
\vspace{-5pt}
 \caption{\label{fig:diffnu1} 
  \footnotesize
 (a) Higher-order relativistic correction, $\Delta g_{s-{\rm nonrec}}^{{\rm rel}(4+)}$,  and (b) three-body correction, $\Delta g_{s-{\rm 3body}}^{(2)}$, as a function of $\nu$ for RV states of H$_2^+$.}
 \end{figure}

Although not required for computing the $g$ factor, it is interesting to calculate the non-adiabatic and recoil contribution of order $\alpha^2$, $\Delta g_{s-{\rm 3-body}}^{(2)}$, by subtracting the non-recoil adiabatic value from the full three-body contribution, according to Eq.~(\ref{eq:gs3b}). Its values for H$_2^+$ RV states are shown in Table~\ref{tab:gs3b} and plotted in Fig.~\ref{fig:diffnu1} (b). They are compatible with the order of magnitude $\sim (m_{\rm e}/m_{\rm p}) \Delta g_s^{(2)}$, and increase with $\nu$ as expected.

%%%%%%%%%%%%%%%%%%%%%%%%%%%%%%%%%%%%%%%%%%%%%%%%%%%%%
\begin{table}[htpb]
\begin{tabular}{cccccc}
\hline
 & \text{$N=0$} & \text{$N=1$} & \text{$N=2$} & \text{$N=3$} & \text{$N=4$} \\
\hline
 \text{$\nu $=0} & 0.98 & 0.99 & 0.99 & 0.99 & 0.99 \\
 \text{$\nu $=1} & 1.05 & 1.05 & 1.05 & 1.05 & 1.06 \\
 \text{$\nu $=2} & 1.13 & 1.13 & 1.13 & 1.14 & 1.14 \\
 \text{$\nu $=3} & 1.23 & 1.23 & 1.23 & 1.24 & 1.24 \\
 \text{$\nu $=4} & 1.35 & 1.35 & 1.35 & 1.35 & 1.36 \\
 \text{$\nu $=5} & 1.48 & 1.48 & 1.48 & 1.48 & 1.49 \\
 \text{$\nu $=6} & 1.62 & 1.62 & 1.62 & 1.63 & 1.63 \\
 \text{$\nu $=7} & 1.76 & 1.77 & 1.77 & 1.77 & 1.78 \\
 \text{$\nu $=8} & 1.92 & 1.92 & 1.92 & 1.93 & 1.93 \\
 \text{$\nu $=9} & 2.07 & 2.07 & 2.07 & 2.08 & 2.08 \\
 \text{$\nu $=10} & 2.21 & 2.21 & 2.22 & 2.22 & 2.23 \\
\hline
\end{tabular} \label{tab:gs3b}
\caption{Values of $\Delta g_{s-{\rm 3-body}}^{(2)}$ [Eq.~(\ref{eq:gs3b})] for RV states of H$_2^+$, in units of $10^{-8}$.}
\end{table}
%%%%%%%%%%%%%%%%%%%%%%%%%%%%%%%%%%%%%%%%%%%%%%%%%%%%%

%%%%%%%%%%%%%%%%%%%%%%%%%%%%%%%%%%%%%%%%%%%%%%%%%%%%%%
%%%%%%%%%%%%%%%%%%%%%%%%%%%%%%%%%%%%%%%%%%%%%%%%%%%%%%%%

%%%%%%%%%%%%%%%%%%%%%%%%%%%%%%%%%%%%%%%%
\subsection{Theoretical uncertainty}
%%%%%%%%%%%%%%%%%%%%%%%%%%%%%%%%%%%%%%%%

The theoretical uncertainty of the $g_s$ values given in Tables~\ref{tab:final-g-H2+} and \ref{tab:final-g-HD+} is dominated by two yet unevaluated QED contributions. The largest one is the nonlogarithmic term of the one-loop self-energy contribution of order $\alpha (Z\alpha)^4$, which we estimated using the 1S state hydrogenic atom result, see Eq.~(\ref{eq:g5SE}). We conservatively estimate the uncertainty of the $\alpha^5$-order correction as equal to 100\% of this estimated term. The second important source of uncertainty is the uncalculated radiative-recoil correction of order $\alpha(Z\alpha)^2(m/M)$, to which we associate an uncertainty equal to $2(m_{\rm e}/m_{\rm p})g_{s-{\rm nonrec}}^{(3)}$. Other sources of uncertainties, such as numerical uncertainties of the calculated terms, higher-order QED corrections, and finite-nuclear-size corrections, are much smaller and can be neglected.

The uncertainty of the $g_t$ component originates from the same terms. The logarithmic part of of the $\alpha (Z\alpha)^4$ self-energy contribution is isotropic, being associated with a contact interaction, but the nonlogarithmic term is expected to have a nonzero tensor component. Considering that the leading-order $\alpha(Z\alpha)^2$ radiative correction to $g_t$ is smaller than the respective correction to $g_s$ by factors of 10 to 15, we assume the same scaling for the $\alpha (Z\alpha)^4$ correction. The uncertainty due to the uncalculated $\alpha(Z\alpha)^2(m/M)$ radiative-recoil correction is estimated to $2(m_{\rm e}/m_{\rm p})g_{t-{\rm nonrec}}^{(3)}$, as done for the scalar component. Overall, this leads to an uncertainty of 1 on the last printed digit of the final $g_t$ values.

%%%%%%%%%%%%%%%%%%%%%%%%%%%%%%%%%%%%%%%%%%%%%%%%%%%%%%%%
\begin{table}[htpb]
\begin{tabular}{cc@{\hskip 0.2in}ccccl@{\hskip 0.3in}cccc}
\hline \hline \\ [-2.0ex]
 $N$ & $\nu$  & $\Delta g_s^{(2)}$ & $\Delta g_{s-{\rm nonrec}}^{(3)}$ & $\Delta g_{s-{\rm nonrec}}^{{\rm rel}(4+)}$ & $\Delta g_{s-{\rm nonrec}}^{(5)}$ & \multicolumn{1}{c}{$g_s$} & $\Delta g_t^{(2)}$ & $\Delta g_{t-{\rm nonrec}}^{(3)}$ & $\Delta g_{t-{\rm nonrec}}^{{\rm rel}(4+)}$ & $g_t$ \\
     &        & $\times 10^5$      & $\times 10^8$      & $\times 10^{10}$                  & $\times 10^{10}$ & & $\times 10^6$      & $\times 10^9$      & $\times 10^{11}$ & $\times 10^6$ \\
\hline \\ [-2.0ex]
 0 & 0 & -4.0731545 & 4.8966 & -5.14 & 2.03(88) & 2.002\,278\,621\,47(10) & - & - & - & 0.000\,00 \\
 0 & 1 & -3.9653315 & 4.7758 & -4.95 & 1.98(85) & 2.002\,279\,698\,51(10) & - & - & - & 0.000\,00 \\
 0 & 2 & -3.8662998 & 4.6644 & -4.79 & 1.93(83) & 2.002\,280\,687\,72(10) & - & - & - & 0.000\,00 \\
 0 & 3 & -3.7757987 & 4.5621 & -4.65 & 1.88(81) & 2.002\,281\,591\,72(10) & - & - & - & 0.000\,00 \\
 0 & 4 & -3.6936417 & 4.4685 & -4.54 & 1.84(79) & 2.002\,282\,412\,36(9) & - & - & - & 0.000\,00 \\
 0 & 5 & -3.6197170 & 4.3835 & -4.45 & 1.80(78) & 2.002\,283\,150\,76(9) & - & - & - & 0.000\,00 \\
 0 & 6 & -3.5539891 & 4.3069 & -4.39 & 1.77(76) & 2.002\,283\,807\,28(9) & - & - & - & 0.000\,00 \\
 0 & 7 & -3.4965021 & 4.2385 & -4.35 & 1.73(75) & 2.002\,284\,381\,47(9) & - & - & - & 0.000\,00 \\
 0 & 8 & -3.4473852 & 4.1784 & -4.33 & 1.70(74) & 2.002\,284\,872\,03(9) & - & - & - & 0.000\,00 \\
 0 & 9 & -3.4068602 & 4.1267 & -4.34 & 1.68(72) & 2.002\,285\,276\,76(9) & - & - & - & 0.000\,00 \\
 0 & 10 & -3.3752512 & 4.0835 & -4.38 & 1.65(71) & 2.002\,285\,592\,41(8) & - & - & - & 0.000\,00 \\
\hline \\ [-2.0ex]
1 & 0 & -4.0692868 & 4.8923 & -5.12 & 2.03(88) & 2.002\,278\,660\,11(10) & -1.059808 & -4.649 & -0.6 & -1.064\,46 \\
 1 & 1 & -3.9617235 & 4.7717 & -4.94 & 1.97(85) & 2.002\,279\,734\,55(10) & -0.988985 & -4.537 & -0.5 & -0.993\,53 \\
 1 & 2 & -3.8629431 & 4.6606 & -4.78 & 1.92(83) & 2.002\,280\,721\,25(10) & -0.919454 & -4.418 & -0.4 & -0.923\,88 \\
 1 & 3 & -3.7726866 & 4.5586 & -4.64 & 1.88(81) & 2.002\,281\,622\,81(10) & -0.851291 & -4.291 & -0.3 & -0.855\,58 \\
 1 & 4 & -3.6907696 & 4.4653 & -4.53 & 1.84(79) & 2.002\,282\,441\,05(9) & -0.784578 & -4.156 & -0.2 & -0.788\,74 \\
 1 & 5 & -3.6170824 & 4.3805 & -4.44 & 1.80(78) & 2.002\,283\,177\,08(9) & -0.719410 & -4.012 & -0.1 & -0.723\,42 \\
 1 & 6 & -3.5515917 & 4.3041 & -4.38 & 1.76(76) & 2.002\,283\,831\,23(9) & -0.655889 & -3.857 & -0.0 & -0.659\,75 \\
 1 & 7 & -3.4943440 & 4.2360 & -4.34 & 1.73(75) & 2.002\,284\,403\,02(9) & -0.594126 & -3.691 & 0.1 & -0.597\,82 \\
 1 & 8 & -3.4454713 & 4.1762 & -4.33 & 1.70(74) & 2.002\,284\,891\,15(9) & -0.534242 & -3.510 & 0.2 & -0.537\,75 \\
 1 & 9 & -3.4051983 & 4.1247 & -4.34 & 1.68(72) & 2.002\,285\,293\,36(9) & -0.476359 & -3.314 & 0.2 & -0.479\,67 \\
 1 & 10 & -3.3738529 & 4.0817 & -4.38 & 1.65(71) & 2.002\,285\,606\,38(8) & -0.420596 & -3.100 & 0.3 & -0.423\,69 \\

\hline \\ [-2.0ex]
 2 & 0 & -4.0615976 & 4.8836 & -5.10 & 2.02(87) & 2.002\,278\,736\,92(10) & -0.893192 & -3.926 & -0.5 & -0.897\,12 \\
 2 & 1 & -3.9545518 & 4.7637 & -4.91 & 1.97(85) & 2.002\,279\,806\,19(10) & -0.833351 & -3.831 & -0.4 & -0.837\,19 \\
 2 & 2 & -3.8562717 & 4.6531 & -4.75 & 1.92(83) & 2.002\,280\,787\,89(10) & -0.774606 & -3.730 & -0.3 & -0.778\,34 \\
 2 & 3 & -3.7665025 & 4.5516 & -4.62 & 1.88(81) & 2.002\,281\,684\,58(10) & -0.717020 & -3.623 & -0.3 & -0.720\,65 \\
 2 & 4 & -3.6850637 & 4.4588 & -4.51 & 1.83(79) & 2.002\,282\,498\,05(9) & -0.660666 & -3.509 & -0.2 & -0.664\,18 \\
 2 & 5 & -3.6118500 & 4.3746 & -4.43 & 1.80(78) & 2.002\,283\,229\,4(9) & -0.605620 & -3.386 & -0.1 & -0.609\,01 \\
 2 & 6 & -3.5468323 & 4.2987 & -4.37 & 1.76(76) & 2.002\,283\,878\,77(9) & -0.551971 & -3.255 & -0.0 & -0.555\,23 \\
 2 & 7 & -3.4900621 & 4.2310 & -4.33 & 1.73(75) & 2.002\,284\,445\,79(9) & -0.499813 & -3.114 & 0.1 & -0.502\,93 \\
 2 & 8 & -3.4416767 & 4.1717 & -4.32 & 1.70(73) & 2.002\,284\,929\,05(9) & -0.449247 & -2.961 & 0.1 & -0.452\,21 \\
 2 & 9 & -3.4019069 & 4.1207 & -4.33 & 1.67(72) & 2.002\,285\,326\,23(9) & -0.400377 & -2.794 & 0.2 & -0.403\,17 \\
 2 & 10 & -3.3710884 & 4.0783 & -4.38 & 1.65(71) & 2.002\,285\,633\,99(8) & -0.353302 & -2.612 & 0.3 & -0.355\,91 \\

\hline \\ [-2.0ex]
 3 & 0 & -4.0501779 & 4.8708 & -5.05 & 2.01(87) & 2.002\,278\,850\,99(10) & -0.859282 & -3.789 & -0.5 & -0.863\,08 \\
 3 & 1 & -3.9439028 & 4.7517 & -4.87 & 1.96(85) & 2.002\,279\,912\,56(10) & -0.801493 & -3.697 & -0.4 & -0.805\,19 \\
 3 & 2 & -3.8463682 & 4.6420 & -4.72 & 1.91(83) & 2.002\,280\,886\,82(10) & -0.744768 & -3.599 & -0.3 & -0.748\,37 \\
 3 & 3 & -3.7573254 & 4.5412 & -4.59 & 1.87(81) & 2.002\,281\,776\,25(9) & -0.689169 & -3.495 & -0.2 & -0.692\,67 \\
 3 & 4 & -3.6765998 & 4.4492 & -4.49 & 1.83(79) & 2.002\,282\,582\,59(9) & -0.634765 & -3.384 & -0.2 & -0.638\,15 \\
 3 & 5 & -3.6040924 & 4.3657 & -4.40 & 1.79(77) & 2.002\,283\,306\,83(9) & -0.581633 & -3.265 & -0.1 & -0.584\,90 \\
 3 & 6 & -3.5397808 & 4.2905 & -4.35 & 1.76(76) & 2.002\,283\,949\,20(9) & -0.529856 & -3.138 & 0.0 & -0.532\,99 \\
 3 & 7 & -3.4837238 & 4.2237 & -4.32 & 1.72(75) & 2.002\,284\,509\,10(9) & -0.479525 & -3.000 & 0.1 & -0.482\,52 \\
 3 & 8 & -3.4360667 & 4.1651 & -4.31 & 1.70(73) & 2.002\,284\,985\,09(9) & -0.430740 & -2.851 & 0.1 & -0.433\,59 \\
 3 & 9 & -3.3970500 & 4.1148 & -4.33 & 1.67(72) & 2.002\,285\,374\,75(8) & -0.383599 & -2.689 & 0.2 & -0.386\,29 \\
 3 & 10 & -3.3670205 & 4.0732 & -4.37 & 1.65(71) & 2.002\,285\,674\,62(8) & -0.338197 & -2.512 & 0.3 & -0.340\,71 \\

\hline \\ [-2.0ex]
 4 & 0 & -4.0351603 & 4.8540 & -5.00 & 2.01(87) & 2.002\,279\,001\,00(10) & -0.843304 & -3.735 & -0.5 & -0.847\,04 \\
 4 & 1 & -3.9299029 & 4.7360 & -4.82 & 1.95(84) & 2.002\,280\,052\,41(10) & -0.786301 & -3.643 & -0.4 & -0.789\,95 \\
 4 & 2 & -3.8333532 & 4.6273 & -4.67 & 1.91(82) & 2.002\,281\,016\,83(10) & -0.730355 & -3.546 & -0.3 & -0.733\,90 \\
 4 & 3 & -3.7452703 & 4.5276 & -4.55 & 1.86(80) & 2.002\,281\,896\,67(9) & -0.675529 & -3.442 & -0.2 & -0.678\,97 \\
 4 & 4 & -3.6654879 & 4.4366 & -4.45 & 1.82(79) & 2.002\,282\,693\,59(9) & -0.621890 & -3.332 & -0.1 & -0.625\,22 \\
 4 & 5 & -3.5939152 & 4.3541 & -4.37 & 1.78(77) & 2.002\,283\,408\,49(9) & -0.569513 & -3.214 & -0.1 & -0.572\,73 \\
 4 & 6 & -3.5305388 & 4.2799 & -4.32 & 1.75(76) & 2.002\,284\,041\,52(9) & -0.518483 & -3.087 & 0.0 & -0.521\,57 \\
 4 & 7 & -3.4754272 & 4.2140 & -4.29 & 1.72(74) & 2.002\,284\,591\,97(9) & -0.468889 & -2.950 & 0.1 & -0.471\,84 \\
 4 & 8 & -3.4287368 & 4.1564 & -4.29 & 1.69(73) & 2.002\,285\,058\,30(8) & -0.420828 & -2.802 & 0.2 & -0.423\,63 \\
 4 & 9 & -3.3907209 & 4.1072 & -4.31 & 1.67(72) & 2.002\,285\,437\,96(8) & -0.374398 & -2.641 & 0.2 & -0.377\,04 \\
 4 & 10 & -3.3617417 & 4.0665 & -4.36 & 1.64(71) & 2.002\,285\,727\,34(8) & -0.329690 & -2.464 & 0.3 & -0.332\,15 \\

\hline \hline \\ [-2.0ex]
\end{tabular}
\caption{Contributions to the scalar and tensor components of the $g$ factor for RV states of H$_2^+$ (see text for details). Our final theoretical predictions for $g_s$ and $g_t$ are given in columns 7 and 11, respectively.} 
\label{tab:final-g-H2+}
\end{table}
%%%%%%%%%%%%%%%%%%%%%%%%%%%%%%%%%%%%%%%%%%%%%%%%%%%%%%%%

%%%%%%%%%%%%%%%%%%%%%%%%%%%%%%%%%%%%%%%%%%%%%%%%%%%%%%%%
\begin{table}[htpb]
\begin{tabular}{cc@{\hskip 0.2in}ccccl@{\hskip 0.3in}cccc}
\hline \hline \\ [-2.0ex]
 $N$ & $\nu$  & $\Delta g_s^{(2)}$ & $\Delta g_{s-{\rm nonrec}}^{(3)}$ & $\Delta g_{s-{\rm nonrec}}^{{\rm rel}(4+)}$ & $\Delta g_{s-{\rm nonrec}}^{(5)}$ & \multicolumn{1}{c}{$g_s$} & $\Delta g_t^{(2)}$ & $\Delta g_{t-{\rm nonrec}}^{(3)}$ & $\Delta g_{t-{\rm nonrec}}^{{\rm rel}(4+)}$ & $g_t$ \\
     &        & $\times 10^5$      & $\times 10^8$      & $\times 10^{10}$                  & $\times 10^{10}$ & & $\times 10^6$      & $\times 10^9$      & $\times 10^{11}$ & $\times 10^6$ \\
\hline \\ [-2.0ex]
 0 & 0 & -4.0812407 & 4.9053 & -5.15 & 2.03(88) & 2.002\,278\,540\,70(10) & - & - & - & 0.000\,00 \\
 0 & 1 & -3.9868059 & 4.7996 & -4.99 & 1.99(86) & 2.002\,279\,484\,00(10) & - & - & - & 0.000\,00 \\
 0 & 2 & -3.8990129 & 4.7009 & -4.84 & 1.94(84) & 2.002\,280\,360\,95(10) & - & - & - & 0.000\,00 \\
 0 & 3 & -3.8176792 & 4.6092 & -4.71 & 1.90(82) & 2.002\,281\,173\,38(10) & - & - & - & 0.000\,00 \\
 0 & 4 & -3.7426643 & 4.5241 & -4.61 & 1.86(81) & 2.002\,281\,922\,69(9) & - & - & - & 0.000\,00 \\
 0 & 5 & -3.6738695 & 4.4455 & -4.52 & 1.83(79) & 2.002\,282\,609\,85(9) & - & - & - & 0.000\,00 \\
 0 & 6 & -3.6112388 & 4.3733 & -4.44 & 1.80(78) & 2.002\,283\,235\,44(9) & - & - & - & 0.000\,00 \\
 0 & 7 & -3.5547577 & 4.3074 & -4.39 & 1.77(76) & 2.002\,283\,799\,60(9) & - & - & - & 0.000\,00 \\
 0 & 8 & -3.5044597 & 4.2477 & -4.35 & 1.74(75) & 2.002\,284\,301\,98(9) & - & - & - & 0.000\,00 \\
 0 & 9 & -3.4604240 & 4.1942 & -4.34 & 1.71(74) & 2.002\,284\,741\,80(9) & - & - & - & 0.000\,00 \\
 0 & 10 & -3.4227819 & 4.1469 & -4.34 & 1.69(73) & 2.002\,285\,117\,75(9) & - & - & - & 0.000\,00 \\
\hline \\ [-2.0ex]
 1 & 0 & -4.0783242 & 4.9021 & -5.14 & 2.03(88) & 2.002\,278\,569\,83(10) & -1.064002 & -4.657 & -0.6 & -1.068\,66 \\
 1 & 1 & -3.9840591 & 4.7965 & -4.98 & 1.98(86) & 2.002\,279\,511\,44(10) & -1.002537 & -4.560 & -0.5 & -1.007\,10 \\
 1 & 2 & -3.8964308 & 4.6980 & -4.83 & 1.94(84) & 2.002\,280\,386\,75(10) & -0.942031 & -4.459 & -0.5 & -0.946\,49 \\
 1 & 3 & -3.8152577 & 4.6064 & -4.71 & 1.90(82) & 2.002\,281\,197\,57(10) & -0.882532 & -4.352 & -0.4 & -0.886\,89 \\
 1 & 4 & -3.7404002 & 4.5215 & -4.60 & 1.86(81) & 2.002\,281\,945\,30(9) & -0.824090 & -4.240 & -0.3 & -0.828\,33 \\
 1 & 5 & -3.6717607 & 4.4431 & -4.51 & 1.83(79) & 2.002\,282\,630\,92(9) & -0.766765 & -4.120 & -0.2 & -0.770\,89 \\
 1 & 6 & -3.6092838 & 4.3711 & -4.44 & 1.79(78) & 2.002\,283\,254\,97(9) & -0.710618 & -3.994 & -0.1 & -0.714\,61 \\
 1 & 7 & -3.5529574 & 4.3053 & -4.38 & 1.76(76) & 2.002\,283\,817\,58(9) & -0.655717 & -3.859 & -0.0 & -0.659\,58 \\
 1 & 8 & -3.5028148 & 4.2458 & -4.35 & 1.74(75) & 2.002\,284\,318\,41(9) & -0.602135 & -3.716 & 0.1 & -0.605\,85 \\
 1 & 9 & -3.4589370 & 4.1924 & -4.33 & 1.71(74) & 2.002\,284\,756\,65(9) & -0.549951 & -3.562 & 0.1 & -0.553\,51 \\
 1 & 10 & -3.4214567 & 4.1453 & -4.34 & 1.69(73) & 2.002\,285\,130\,98(9) & -0.499244 & -3.397 & 0.2 & -0.502\,64 \\
\hline \\ [-2.0ex]
  2 & 0 & -4.0725173 & 4.8956 & -5.12 & 2.03(88) & 2.002\,278\,627\,84(10) & -0.897364 & -3.934 & -0.5 & -0.901\,30 \\
 2 & 1 & -3.9785906 & 4.7903 & -4.96 & 1.98(86) & 2.002\,279\,566\,06(10) & -0.845427 & -3.852 & -0.5 & -0.849\,28 \\
 2 & 2 & -3.8912906 & 4.6922 & -4.81 & 1.94(84) & 2.002\,280\,438\,09(10) & -0.794302 & -3.766 & -0.4 & -0.798\,07 \\
 2 & 3 & -3.8104377 & 4.6010 & -4.69 & 1.90(82) & 2.002\,281\,245\,72(10) & -0.744029 & -3.676 & -0.3 & -0.747\,71 \\
 2 & 4 & -3.7358942 & 4.5164 & -4.58 & 1.86(80) & 2.002\,281\,990\,31(9) & -0.694654 & -3.580 & -0.2 & -0.698\,24 \\
 2 & 5 & -3.6675646 & 4.4384 & -4.50 & 1.82(79) & 2.002\,282\,672\,83(9) & -0.646224 & -3.479 & -0.2 & -0.649\,70 \\
 2 & 6 & -3.6053954 & 4.3667 & -4.43 & 1.79(77) & 2.002\,283\,293\,81(9) & -0.598792 & -3.372 & -0.1 & -0.602\,17 \\
 2 & 7 & -3.5493768 & 4.3012 & -4.37 & 1.76(76) & 2.002\,283\,853\,35(9) & -0.552416 & -3.258 & -0.0 & -0.555\,67 \\
 2 & 8 & -3.4995444 & 4.2420 & -4.34 & 1.73(75) & 2.002\,284\,351\,08(9) & -0.507158 & -3.136 & 0.1 & -0.510\,29 \\
 2 & 9 & -3.4559819 & 4.1890 & -4.33 & 1.71(74) & 2.002\,284\,786\,17(9) & -0.463084 & -3.006 & 0.1 & -0.466\,09 \\
 2 & 10 & -3.4188248 & 4.1422 & -4.33 & 1.69(73) & 2.002\,285\,157\,27(9) & -0.420260 & -2.866 & 0.2 & -0.423\,12 \\
\hline \\ [-2.0ex]
 3 & 0 & -4.0638718 & 4.8859 & -5.09 & 2.02(87) & 2.002\,278\,714\,20(10) & -0.864217 & -3.797 & -0.5 & -0.868\,02 \\
 3 & 1 & -3.9704500 & 4.7812 & -4.93 & 1.98(85) & 2.002\,279\,647\,38(10) & -0.814054 & -3.718 & -0.4 & -0.817\,78 \\
 3 & 2 & -3.8836400 & 4.6836 & -4.79 & 1.93(84) & 2.002\,280\,514\,51(10) & -0.764680 & -3.635 & -0.4 & -0.768\,32 \\
 3 & 3 & -3.8032650 & 4.5929 & -4.66 & 1.89(82) & 2.002\,281\,317\,36(10) & -0.716134 & -3.547 & -0.3 & -0.719\,68 \\
 3 & 4 & -3.7291904 & 4.5089 & -4.56 & 1.86(80) & 2.002\,282\,057\,28(9) & -0.668457 & -3.455 & -0.2 & -0.671\,91 \\
 3 & 5 & -3.6613235 & 4.4313 & -4.48 & 1.82(79) & 2.002\,282\,735\,18(9) & -0.621697 & -3.357 & -0.1 & -0.625\,05 \\
 3 & 6 & -3.5996140 & 4.3601 & -4.41 & 1.79(77) & 2.002\,283\,351\,56(9) & -0.575904 & -3.253 & -0.1 & -0.579\,16 \\
 3 & 7 & -3.5440554 & 4.2951 & -4.36 & 1.76(76) & 2.002\,283\,906\,50(9) & -0.531135 & -3.142 & -0.0 & -0.534\,28 \\
 3 & 8 & -3.4946869 & 4.2363 & -4.33 & 1.73(75) & 2.002\,284\,399\,60(9) & -0.487450 & -3.024 & 0.1 & -0.490\,47 \\
 3 & 9 & -3.4515960 & 4.1838 & -4.31 & 1.71(74) & 2.002\,284\,829\,98(9) & -0.444912 & -2.898 & 0.1 & -0.447\,81 \\
 3 & 10 & -3.4149228 & 4.1375 & -4.32 & 1.68(73) & 2.002\,285\,196\,25(9) & -0.403586 & -2.762 & 0.2 & -0.406\,35 \\
\hline \\ [-2.0ex]
 4 & 0 & -4.0524634 & 4.8731 & -5.05 & 2.02(87) & 2.002\,278\,828\,16(10) & -0.849353 & -3.744 & -0.5 & -0.853\,10 \\
 4 & 1 & -3.9597100 & 4.7691 & -4.89 & 1.97(85) & 2.002\,279\,754\,66(10) & -0.799866 & -3.665 & -0.4 & -0.803\,54 \\
 4 & 2 & -3.8735487 & 4.6723 & -4.75 & 1.93(83) & 2.002\,280\,615\,32(10) & -0.751161 & -3.583 & -0.3 & -0.754\,75 \\
 4 & 3 & -3.7938067 & 4.5823 & -4.63 & 1.89(82) & 2.002\,281\,411\,84(10) & -0.703276 & -3.496 & -0.3 & -0.706\,77 \\
 4 & 4 & -3.7203532 & 4.4989 & -4.53 & 1.85(80) & 2.002\,282\,145\,55(9) & -0.656255 & -3.404 & -0.2 & -0.659\,66 \\
 4 & 5 & -3.6530995 & 4.4219 & -4.45 & 1.82(78) & 2.002\,282\,817\,32(9) & -0.610142 & -3.307 & -0.1 & -0.613\,45 \\
 4 & 6 & -3.5919996 & 4.3514 & -4.39 & 1.78(77) & 2.002\,283\,427\,62(9) & -0.564990 & -3.204 & -0.1 & -0.568\,19 \\
 4 & 7 & -3.5370513 & 4.2870 & -4.34 & 1.75(76) & 2.002\,283\,976\,46(9) & -0.520852 & -3.094 & 0.0 & -0.523\,95 \\
 4 & 8 & -3.4882985 & 4.2289 & -4.31 & 1.73(75) & 2.002\,284\,463\,41(9) & -0.477789 & -2.977 & 0.1 & -0.480\,76 \\
 4 & 9 & -3.4458341 & 4.1770 & -4.30 & 1.70(74) & 2.002\,284\,887\,53(9) & -0.435863 & -2.851 & 0.1 & -0.438\,71 \\
 4 & 10 & -3.4098041 & 4.1314 & -4.31 & 1.68(73) & 2.002\,285\,247\,37(9) & -0.395137 & -2.716 & 0.2 & -0.397\,85 \\
\hline \hline \\ [-2.0ex]
\end{tabular}
\caption{Same as Table~\ref{tab:final-g-H2+}, for the RV states of HD$^+$.} \label{tab:final-g-HD+}
\end{table}
%%%%%%%%%%%%%%%%%%%%%%%%%%%%%%%%%%%%%%%%%%%%%%%%%%%%%%%%

%%%%%%%%%%%%%%%%%%%%%%%%
\section{Conclusion}
%%%%%%%%%%%%%%%%%%%%%%%%

We have presented calculations of relativistic and QED corrections to the bound-electron $g$ factor in the MHI H$_2^+$ and HD$^+$. The theoretical precision is improved by a factor of approximately 2000, from $10^{-7}$ to $4-5\times 10^{-11}$. These results allow for greatly improved predictions of spin-flip frequencies in a large magnetic field, which will facilitate internal state identification in single-ion Penning trap experiments~\cite{Myers2018,Koenig2025}. The gain of precision is also useful for ro-vibrational spectroscopy of single MHI in Penning traps, as it allows for accurate predictions of the Zeeman shift. For example, in a 4~T magnetic field, the uncertainty of the Zeeman shift of an individual RV state due to the bound-electron $g$ factor is now reduced to about 5~Hz.

Finally, these results pave the way towards highly accurate comparisons between experimental and theoretical $g$ factors. The experimental advances demonstrated in~\cite{Koenig2025} allow for a high-precision measurement in HD$^+$, which has been recently performed and will be compared to theoretical predictions in a forthcoming paper.

On the theoretical side, there is strong potential to improve the precision to very high levels since the NRQED expansion is most efficient for $Z=1$. For example, the $g$ factor of the $^3$He$^+$ ion ($Z=2$) has been calculated with a relative uncertainty of $1.9 \times 10^{-13}$~\cite{Schneider2022}. Comparable accuracy could be reached in MHI, which on the one hand are more complex, but on the other hand have the favorable feature of a lower nuclear charge. In that regard, it is worth noting that the current relative theoretical uncertainty of RV transition frequencies in HMI ($\sim 8 \times 10^{-12}$)~\cite{Korobov2021} is close to that of the 1S-2S transition frequency in He$^+$ ($\sim 4 \times 10^{-12}$)~\cite{Moreno2023}.

Reaching a similar accuracy level in $g$ factor calculations as achieved in He$^+$ would open new avenues to fundamental physics tests thanks to the possibility of measuring $g$ factor differences with very high accuracy, as demonstrated in~\cite{Sailer2022}. For example, a high-precision test of the CPT symmetry has been proposed using a $g$ factor comparison between a free positron and a He$^+$ ion~\cite{lsym}. A similar comparison between a positron and a MHI could be envisaged. In addition, a He$^+$/HMI comparison would allow for a stringent test of QED.

%%%%%%%%%%%%%%%%%%%%%%%%%%%%%
\section*{Acknowledgments}
%%%%%%%%%%%%%%%%%%%%%%%%%%%%%
We are grateful to S. Schiller (Heinrich-Heine-Universit\"at D\"usseldorf) for motivating the calculation of the Dirac $g$ factor. We also thank him, as well as F. Hei\ss{}e, C. K\"onig and S. Sturm (MPIK Heidelberg) for helpful discussions and kind support. O. K. thanks M. Garcia (Universit{\"a}t Kassel) for his kind support. We  thank the computing centers at Universit{\"a}t Kassel and Universit{\"a}t D{\"u}sseldorf for providing resources and advice. The work of H.N. and J.-Ph.K. was part of 23FUN04 COMOMET that has received funding from the European Partnership on Metrology, co-financed by the European Union's Horizon Europe Research and Innovation Programme and from by the Participating States. Funder ID: 10.13039/100019599.

%%%%%%%%%%%%%%%%%%%%%%%%%%%%%
\appendix
%%%%%%%%%%%%%%%%%%%%%%%%%%%%%
\section{}\label{apA}
%%%%%%%%%%%%%%%%%%%%%%%%%%%%%

In this Appendix, we give additional details on the derivation of the components of the $g$ tensor.

%%%%%%%%%%%%%%%%%%%%%%%%%%%%%%%%%%%%%%%%%%%%%%%%%%%%%%%%%%%%%%%%%
\subsection{Parallel case}\label{apA1}
%%%%%%%%%%%%%%%%%%%%%%%%%%%%%%%%%%%%%%%%%%%%%%%%%%%%%%%%%%%%%%%%%

Here, we detail the derivation of Eq.~\ref{gpar}. Using Eq. \ref{PaMat1} (with $f_{\pm}=\rho\,e^{\pm i\phi}$), we get
\begin{eqnarray}
&&\langle\Psi_{+}\mid M_{||}(\rho,\phi) \mid\Psi_{+}\rangle\\\nonumber 
&=&
\begin{bmatrix}
 \psi_1& \\
 \psi_2\, e^{-i \phi}& \\
 -i \psi_3&\\
 -i \psi_4\, e^{-i \phi}
 \end{bmatrix}
 \begin{bmatrix} 
  0 & 0 & 0&\,  -i \, f_{-}\\ 
  0 & 0 & i \,  f_{+}&0 \\
  0& -i \, f_{-} & 0&0 \\
  i \, f_{+}&0 & 0&0 
\end{bmatrix}
\begin{bmatrix}
 \psi_1& \\
 \psi_2\, e^{i \phi}& \\
 i \psi_3&\\
 i \psi_4\, e^{i \phi}
 \end{bmatrix}
 \\\nonumber
 &=&
 \begin{bmatrix}
 \psi_1& \\
 \psi_2\, e^{-i \phi}& \\
 -i \psi_3&\\
 -i \psi_4\, e^{-i \phi}
 \end{bmatrix}
 \begin{bmatrix}
 \rho \psi_4& \\
 -\rho \psi_3\, e^{i \phi}& \\
 -i \rho \psi_2&\\
 i \rho \psi_1\, e^{i \phi}
 \end{bmatrix}
 \\\nonumber
 &=&
 2\langle\psi_1\mid\rho\mid\psi_4\rangle- 
  2 \langle\psi_2\mid\rho\mid\psi_3\rangle \\
&=& 4 \pi \left( \langle\psi_1\mid\rho\mid\psi_4\rangle_{\rm rad} - 
  \langle\psi_2\mid\rho\mid\psi_3\rangle_{\rm rad} \right), \nonumber
\end{eqnarray}
where $\langle \rangle_{\rm rad}$ denotes a radial integral obtained after integration over $\phi$. It is worth noting that the expression in the line before last is identical to that obtained for one-electron atoms in Eq.~(1) of Ref.~\cite{Margenau1940}.

%%%%%%%%%%%%%%%%%%%%%%%%%%%%%%%%%%%%%%%%%%%%%%%%%%
\subsection{Perpendicular case}\label{apA2}
%%%%%%%%%%%%%%%%%%%%%%%%%%%%%%%%%%%%%%%%%%%%%%%%%%

We detail the derivation of Eqs.~(\ref{gperp1}) and (\ref{gperp2}). Using Eq.~(\ref{Hperpx}) one gets for the first term:
\begin{eqnarray}\label{AppA:perp2}
&&\langle\Psi_{-}\mid M_{\perp}(z)\mid\Psi_{+}\rangle \\\nonumber
&=&
\begin{bmatrix}
 -\psi_2\, e^{i \phi}& \\
  \psi_1& \\
 -i \psi_4\, e^{i \phi}\\ 
  i \psi_3&\\
 \end{bmatrix}
 \begin{bmatrix} 
  0 & 0 &&  -i\,z \\
  0 & 0 & i\,z &\\
  &  -i\,z & 0&0 \\
  i\,z  & &0 & 0
\end{bmatrix}
\begin{bmatrix}
 \psi_1& \\
 \psi_2\, e^{i \phi}& \\
 i \psi_3&\\
 i \psi_4\, e^{i \phi}
 \end{bmatrix}
 \\\nonumber
 &=&
 \begin{bmatrix}
 -\psi_2\, e^{i \phi}& \\
  \psi_1& \\
 -i \psi_4\, e^{i \phi}\\ 
  i \psi_3&\\
 \end{bmatrix}
 \begin{bmatrix}
  z\, \psi_4\, e^{i \phi} & \\
  -z\, \psi_3& \\
 -i\, z\psi_2\, e^{i \phi} \\
  i\, z \psi_1&\\
 \end{bmatrix}
 \\\nonumber
 &=&
  -2 \langle\psi_2 \mid z\, e^{i 2\phi} \mid\psi_4\, \rangle 
 -2 \langle\psi_1 \mid z\mid\psi_3\, \rangle 
 \\\nonumber
 &=&
 -4\pi \langle\psi_1 \mid z\mid\psi_3\, \rangle_{\rm rad}.
\end{eqnarray}
For the second term of Eq.~(\ref{Hperpx}) we have (with $f=\rho\, \sin(\phi))$)
\begin{eqnarray}\label{AppA:perp1}
&&\langle\Psi_{-}\mid M_{\perp}(\rho,\phi)\mid\Psi_{+}\rangle
\\\nonumber &=&
\begin{bmatrix}
 -\psi_2\, e^{i \phi}& \\
  \psi_1& \\
 -i \psi_4\, e^{i \phi}\\ 
  i \psi_3&\\
 \end{bmatrix}
 \begin{bmatrix} 
  0 & 0 &-f&  0 \\
  0 & 0 & 0 &f\\
  -f& 0 & 0&0 \\
  0 & f &0 & 0
\end{bmatrix}
\begin{bmatrix}
 \psi_1& \\
 \psi_2\, e^{i \phi}& \\
 i \psi_3&\\
 i \psi_4\, e^{i \phi}
 \end{bmatrix}
 \\\nonumber
 &=&
 \begin{bmatrix}
 -\psi_2\, e^{i \phi}& \\
  \psi_1& \\
 -i \psi_4\, e^{i \phi}\\ 
  i \psi_3&\\
 \end{bmatrix}
 \begin{bmatrix}
 -i\rho \sin(\phi) \,\psi_3& \\
 i\rho\sin(\phi) \, \psi_4\, e^{i \phi}& \\
 -\rho\sin(\phi) \psi_1&\\
 \rho\sin(\phi) \psi_2\, e^{i \phi}
 \end{bmatrix}
 \\\nonumber
 &=&
 i\, \langle\psi_2\mid\rho\sin(\phi)\, e^{i \phi}\mid\psi_3\rangle 
 +i\, \langle\psi_1\mid\rho\sin(\phi)\, e^{i \phi}\mid\psi_4\rangle
 \\\nonumber
 &&
 +i\, \langle\psi_4\mid\rho\sin(\phi)\, e^{i \phi}\mid\psi_1\rangle 
 +i\, \langle\psi_3\mid\rho\sin(\phi)\, e^{i \phi}\mid\psi_2\rangle
 \\\nonumber
 &=&
  - 2\pi \left( \langle\psi_4\mid\rho\mid\psi_1\rangle_{\rm rad}
 + \langle\psi_2\mid\rho\mid\psi_3\rangle_{\rm rad} \right),
\end{eqnarray}
where in the last step the integration over the angle $\phi$ yields
$i\,\int_{0}^{2\pi}  \sin(\phi)\, e^{i \phi} = -\pi $. 

%%%%%%%%%%%%%%%%%%%%%%%%%%%%%%%%
\section{} \label{app-largeR}
%%%%%%%%%%%%%%%%%%%%%%%%%%%%%%%%

The object of this section is to study the $g$ tensor components of $\rm H_{2}^{+}$ in the large-$R$ limit, as a further test of our calculations. Results as a function of $R$ are given in Table \ref{tab:g_H2+_largeR}.

%%%%%%%%%%%%%%%%%%%%%%%%%%%%%%%%%%%%%%%%%%%%%%%%%%
\begin{table}[thpb]
  \begin{tabular*}{\linewidth}{@{\extracolsep{\fill}} llll} \hline\hline\\[-2.0ex]
 $R$ (a.u.) & \hspace{0.50cm} $g_{||}$\, [Eq.~(\ref{gparallel})] & \hspace{0.30cm} $g_{\perp}$\,\, [Eq.~(\ref{gperp3}), (\ref{gperp4})] & \hspace{0.5cm}{$g$}\, [Eq.~(\ref{gMargenau})]
 \\\hline \\ [-2.00ex]
50    & {\bf 1.999964499}03508654 & {\bf 1.999964498}5397285816 &                       \\
100   & {\bf 1.9999644986}7042536 & {\bf 1.9999644986}088512218 &                       \\
500   & {\bf 1.999964498624}69098  & {\bf 1.999964498624}2005373& {\bf 1.9999644986243}636666  \\
1000  & {\bf 1.9999644986243}9765  & {\bf 1.9999644986243}361505& {\bf 1.999964498624356}7500 \\
1500  & {\bf 1.9999644986243}6820  & {\bf 1.99996449862435}00269& {\bf 1.9999644986243560}967 \\
{\rm diff} & $1.\, 10^{-14}$&  $6.\, 10^{-15}$ & $9.6\,10^{-17}$  
 \\\hline 
\end{tabular*}
\caption{\label{tab:g_H2+_largeR} 
Relativistic $g$ tensor values of $H_2^{+}$ for large values of the internuclear distance $R$. diff denotes the absolute error of the values at $R=1500$~a.u. with respect to the exact H-atom value (see Table  \ref{tab:g_Hatom}). Parameters: $p=10$, $\nu = 4$ ($\nu$ is not chosen too high to avoid lowering the density of grid points in the outer region). The grid point number $N$ and the extension of the domain $D_{max}$ are chosen so that the specified bold digits are converged.}
\end{table}
%%%%%%%%%%%%%%%%%%%%%%%%%%%%%%%%%%%%%%%%%%%%%%%%

The precision is not as good as in table \ref{tab:g_Hatom}, because only a small part of the integration domain, i. e. the regions around both nuclei, gives almost the entire contribution. The $g$ factor of a H-like atom can also be calculated with the equation given in \cite{Margenau1940},
\begin{eqnarray}\label{gMargenau}
   g_{\rm Margenau}=2\left[1-(4/3) (\langle \psi_3\mid\psi_3\rangle+\langle\psi_4\mid\psi_4\rangle) \right] \mbox{ if } j=l+1/2.
\end{eqnarray} 
Although this expression is only valid in the atomic case, it will also converge towards the atomic value in the large-$R$ limit. $g$ factor values calculated using this expression are given in the last column of Table~\ref{tab:g_H2+_largeR} for $R\ge 500$~a.u. They yield more precise results, which is due to the fact that the deviation from the free Dirac value $g = 2$ is directly calculated numerically in Eq.~(\ref{gMargenau}). Going to larger $R$ does not allow to reproduce the atomic result with higher precision (as confirmed by a test for $R=2000$~a.u.), because the system size becomes too large, and the number of grid points is no longer enough to get an adequate distribution near the nuclei and in the outer region.

%%%%%%%%%%%%%%%%%%%%%%%%%%%%%%%%%%%%%%%%%%%%%%%%%%%%%%%%%%%%%%%%%
\bibliography{bibliography}
%%%%%%%%%%%%%%%%%%%%%%%%%%%%%%%%%%%%%%%%%%%%%%%%%%%%%%%%%%%%%%%%
\end{document}